\documentclass[a4paper,10pt]{revtex4}
\usepackage{mathrsfs}
\usepackage{graphicx}
\usepackage{latexsym}
\usepackage{amsmath}
\usepackage{amssymb}
\usepackage{textcomp}
\usepackage{amsbsy}
\usepackage{graphics}
\usepackage{epstopdf}
\usepackage{color}

\begin{document}

\tolerance=5000

\title{Viable non-singular cosmic bounce in holonomy improved F(R) gravity endowed with a Lagrange multiplier}

\author{Emilio~Elizalde,$^{1}$\,\thanks{elizalde@ieec.uab.es}
S.~D.~Odintsov,$^{2,3}$\,\thanks{odintsov@ieec.uab.es}
Tanmoy~Paul$^{4,5}$\thanks{pul.tnmy9@gmail.com}}
\affiliation{ $^{1)}$ Institut de Ci\`encies de l'Espai (ICE-CSIC/IEEC),\\ 
Campus UAB, c. Can Magrans s/n, 08193, Barcelona, Spain. \\
$^{2)}$ ICREA, Passeig Luis Companys, 23, 08010 Barcelona, Spain\\
$^{3)}$ Institute of Space Sciences (IEEC-CSIC) C. Can Magrans
s/n,
08193 Barcelona, Spain\\
$^{4)}$ Department of Physics, Chandernagore College, Hooghly - 712 136.\\
$^{(5)}$ Department of Theoretical Physics,\\
Indian Association for the Cultivation of Science,\\
2A $\&$ 2B Raja S.C. Mullick Road,\\
Kolkata - 700 032, India }


\tolerance=5000

\begin{abstract}
Matter and quasi-matter bounce scenarios are studied for an F(R) gravity model with  holonomy corrections and a Lagrange multiplier, 
with a scale factor $a(t) = \left(a_0t^2 + 1 \right)^n$,  where the Hubble parameter squared has a linear and a quadratic dependence 
on the effective energy density. Provided $n < 1/2$, it is shown that the primordial curvature perturbations are generated deeply into 
the contracting era, at large negative time, which makes the low-curvature limit a good approximation for calculating the perturbation power spectrum. 
Moreover, it is shown that, for $n$ within this range, the obtained cosmological quantities are fully compatible with the Planck constraints, 
and that the ``low curvature limit'' comes as a viable approximation to calculate the power spectra of both scalar and tensor perturbations. 
Using  reconstruction techniques for F(R) gravity with the Lagrange multiplier,  the precise form of the effective F(R) gravity is found, 
from which one determines the power spectra of scalar and tensor perturbations in such bouncing scenario. Correspondingly, the spectral index 
for curvature perturbations and the tensor to scalar ratio are obtained, and  these values are successfully confronted  with the 
latest Planck observations. Further, it is shown that both the weak and the null energy conditions are satisfied, thanks to the holonomy corrections 
performed in the theory--which are then proven to be necessary for achieving this goal. In fact, when approaching the bouncing era, 
the holonomy corrections become significant and play a crucial role in order to restore the energy conditions. Summing up, 
a cosmological bouncing scenario with the scale factor above and fulfilling the energy conditions can be adequately described by the F(R) model 
with a Lagrange multiplier and holonomy corrections, which prove to be very important.
\end{abstract}


\maketitle
\section{Introduction}

There seems to be no doubt that, at present, our universe expansion is accelerating. Its expansion rate is quantified by the Hubble parameter, defined as $H = \dot{a}/a$, where 
$a(t)$ is a scale factor of the universe at cosmic time $t$. When we go back in time, there are namely two possibilities: (i) the scale factor started from a value 
zero, what leads to the divergence of the Kretschmann scalar, which in turn ensures the singularity in the spacetime curvature, known as the Big Bang singularity. 
It may be possible that this singularity is just a manifestation of the shortcomings of classical gravity, unable to describe such small scales (in other words, extremely high energy ones). Quite possibly, a yet-to-be-built quantum theory of gravity will be able to resolve the Big Bang singularity, as turns out to be the case in classical electrodynamics with the Coulomb potential singularities at the origin of the potential, which 
are resolved in the context of quantum electrodynamics. (ii) In the absence of a fully accepted quantum gravity theory, there is at present another possibility to deal with this issue within the domain of classical  gravity. This is known as the bouncing scenario 
\cite{Brandenberger:2012zb,Brandenberger:2016vhg,Battefeld:2014uga,Novello:2008ra,Cai:2014bea,deHaro:2015wda,Lehners:2011kr,Lehners:2008vx,
Cheung:2016wik,Cai:2016hea,Cattoen:2005dx,Li:2014era,Brizuela:2009nk,Cai:2013kja,Quintin:2014oea,Cai:2013vm,Poplawski:2011jz,
Koehn:2015vvy,Odintsov:2015zza,Nojiri:2016ygo,Oikonomou:2015qha,Odintsov:2015ynk,Koehn:2013upa,Battarra:2014kga,Martin:2001ue,Khoury:2001wf,
Buchbinder:2007ad,Brown:2004cs,Hackworth:2004xb,Nojiri:2006ww,Johnson:2011aa,Peter:2002cn,Gasperini:2003pb,Creminelli:2004jg,Lehners:2015mra,
Mielczarek:2010ga,Lehners:2013cka,Cai:2014xxa,Cai:2007qw,Cai:2010zma,Avelino:2012ue,Barrow:2004ad,Haro:2015zda,Elizalde:2014uba,tp}, 
in which the spacetime curvature singularity is absent. In bouncing cosmology, 
the universe starts from a contracting era until it reaches a minimal size, then it bounces off at some specific cosmic time and starts to expand. Thereby, 
the scale factor of the bounce universe does never hit the value zero, which makes the spacetime curvature free from any singularity. Moreover, 
bounce cosmology is also appealing since it can be obtained as a cosmological solution of the theory of Loop Quantum Cosmology 
\cite{Laguna:2006wr,Corichi:2007am,Bojowald:2008pu,Singh:2006im,
Date:2004fj,deHaro:2012xj,Cianfrani:2010ji,Cai:2014zga,
Mielczarek:2008zz,Mielczarek:2008zv,Diener:2014mia,Haro:2015oqa,
Zhang:2011qq,Zhang:2011vi,Cai:2014jla,WilsonEwing:2012pu}.

Among the various bouncing models proposed so far, the matter bounce scenario (MBS) \cite{deHaro:2015wda,Finelli:2001sr,Quintin:2014oea,Cai:2011ci,
Haro:2015zta,Cai:2011zx,Cai:2013kja,
Haro:2014wha,Brandenberger:2009yt,deHaro:2014kxa,Odintsov:2014gea,
Qiu:2010ch,Oikonomou:2014jua,Bamba:2012ka,deHaro:2012xj,
WilsonEwing:2012pu,Cai:2011tc} has earned special attention because it 
generates a nearly scale invariant power spectrum. The MBS is characterized by a universe depicted by a matter dominated epoch at very large negative time 
in the contracting phase, where the primordial curvature perturbations are generated deeply inside the Hubble radius, and is thus able to solve the 
horizon problem. After it bounces off, the universe enters a regular expanding phase (symmetric to the contracting phase), 
in which it matches the behavior of the standard Big Bang cosmology. However, in order to obtain a viable matter bounce scenario, the observable parameters 
of the underlying model have to fulfill a number of stringent constraints, coming from the latest Planck 2018 and other astronomical observations. Along this direction, there 
are still questions to be answered, within the framework of the matter bounce scenario. Firstly, in an exact MBS, characterized by a single scalar field, the power spectrum turns out to be 
exactly scale invariant (i.e the spectral index of the curvature perturbation is exactly equal to one), what is in tension with the observational constraints. Such inconsistency of the spectral index in the context of a matter bounce scenario 
was also confirmed in \cite{Odintsov:2014gea} from a slightly different point of view, namely from an F(R) theory of gravity.  
F(R) models can be equivalently 
mapped to scalar-tensor ones via conformal transformation of the metric \cite{Das:2017htt,Elizalde:2018rmz,Elizalde:2018now} 
and, thus, the inconsistencies of the spectral index from two different models 
are well justified. Secondly, according to the Planck 2018 data, the running of the spectral index is constrained to be $-0.0085 \pm 0.0073$. 
However, for the MBS in the case of a single scalar field model, the running of the index becomes zero and hence it is not compatible with observations. 
Thirdly, in the simplest MBS model the amplitude of scalar fluctuation is found to be comparable to that of tensor perturbations, 
which in turn makes the value of the tensor-to-scalar ratio to be of order one, again in conflict with the Planck constraints. 

However, in a so-called quasi-matter bounce scenario (improving the exact matter bounce one), according to which the scale factor of the 
Universe evolves as $t^{3(1+w)}$ (with $w \neq 0$), deeply in the contracting era it is possible to recover the consistency of the spectral 
index and of the running index, even in a single scalar field model. However, the tensor-to-scalar ratio still remains problematic. Fourthly, 
a crucial drawback of the matter bounce scenario (just as it happens in most of the bouncing models) is the violation of the null energy condition 
by which the bouncing can be realized. 

From a different perspective, it has  been shown \cite{tp_bouncing1}  that 
an F(R) gravity model with Lagrange multiplier is able to resolve most of the problems arising in the context of matter or quasi-matter bounce scenarios, 
albeit it fails to restore the energy conditions.

Motivated by all these arguments, we will here study matter and quasi-matter bounce scenarios in an F(R) gravity model with a Lagrange multiplier 
and with the addition of holonomy corrections, which will be proven to be crucial to solve the remaining problems in the description above.  
The Hubble parameter squared ($H^2$) will be proportional to a linear as well as to a quadratic power of the effective energy 
density ($\rho_\mathrm{eff}$), unlike in the usual Friedmann cosmology, where $H^2$ is proportional  to the linear power of $\rho_\mathrm{eff}$, only. 
We will explore in this 
framework the viability of the matter and quasi-matter bounce scenarios, including, in special, the investigation of the energy conditions.

The paper is organized as follows. In Sects.~[\ref{sec_einstein}], [\ref{sec_F(R)}], and [\ref{sec_LMC}] we discuss how the holonomy corrections modify the Hamiltonian expression for Einstein's gravity, 
F(R) gravity, and the Lagrange multiplier F(R) gravity model, respectively.  Sects.~[\ref{sec_reconstruction}], [\ref{sec_perturbation}], and 
[\ref{sec_energy}] are devoted to the explicit calculation of the power spectrum, the observational indices, and the investigation of the energy
conditions, in the holonomy corrected Lagrange multiplier $F(R)$ gravity model. Conclusions follow at the end of the paper.

A technical observation is here in order. Before considering the holonomy modifications of the  F(R) gravity model with a Lagrange multiplier (an essential ingredient of the present work), it is convenient to discuss the issue of the holonomy 
improvement in the more standard cases of the Einstein and pure F(R) gravity models. It should be mentioned that  holonomy corrections can be introduced 
in a variety of ways, which are connected to one another by a canonical transformation (see \cite{haro1}, for the details). We will here restrict our analysis to a particular procedure for introducing the holonomy improvement (noted by unbarred quantities in \cite{haro1}).

\section{Einstein gravity with a scalar field}\label{sec_einstein}

\subsection{Ordinary case without holonomy corrections}
In the flat Friedmann-Lemaitre-Robertson-Walker (FLRW) spacetime, the Lagrangian of Einstein's gravity along with a scalar field can be written as,
\begin{eqnarray}
 L(V,\dot{V},\ddot{V}) = \frac{1}{2}VR + V\big[\frac{1}{2}\dot{\Phi}^2 - U(\Phi)\big]
 \label{eL1}
 \end{eqnarray}
where $V = a^3$ is the volume, $R = \frac{2\ddot{V}}{V} - \frac{2\dot{V}^2}{3V^2}$ is the scalar curvature and $\Phi$ the scalar field endowed 
with the potential $U(\Phi)$. It should be noticed that the above Lagrangian contains higher derivatives (a second derivative) term of $V$ and 
thus, to get the Hamiltonian from this Lagrangian, it is useful to introduce a Lagrange multiplier, namely $\mu$, as follows 
 \begin{eqnarray}
  L_1(V,\dot{V},\ddot{V},R) = \frac{1}{2}VR + \frac{1}{2}\mu\big[\frac{2\ddot{V}}{V} - \frac{2\dot{V}^2}{3V^2} - R\big] + V\big[\frac{1}{2}\dot{\Phi}^2 - U(\Phi)\big]
  \label{eL2}
 \end{eqnarray}
 The equation of motion for $R$ is given by $\mu = \frac{1}{2}V$. In order to remove the second derivative $\ddot{V}$, we subtract a total derivative term 
 $\frac{d\dot{V}}{dt}$ from the above Lagrangian as (note that the subtraction of a total derivative from a Lagrangian does not change its dynamics)
 \begin{eqnarray}
  \tilde{L}(V,\dot{V},R,\dot{R})&=&L_1 - \frac{d}{dt}\big(\dot{V}\big)\nonumber\\
  &=&-\frac{1}{3}\frac{\dot{V}^2}{V} + V\big[\frac{1}{2}\dot{\Phi}^2 - U(\Phi)\big]
  \label{eL3}
 \end{eqnarray}
where we have used $\mu = V/2$. We may note that $\tilde{L}$ depends on $(V,R)$ and its first derivatives. The corresponding conjugate momenta are 
 \begin{eqnarray}
  p_\mathrm{V}&=&\frac{\partial\tilde{L}}{\partial\dot{V}} = -\frac{2\dot{V}}{3V}\nonumber\\
  p_{\Phi}&=&\frac{\partial\tilde{L}}{\partial\dot{\Phi}} = V\dot{\Phi}
  \nonumber
 \end{eqnarray}
 With these expressions for the momenta along with Eq.~(\ref{eL3}), we get the Hamiltonian for Einstein's gravity with a scalar field, as 
 \begin{eqnarray}
  \tilde{H}(V,R,p_V,p_R)&=&p_V \dot{V} + p_{\Phi}\dot{\Phi} - \tilde{L}\nonumber\\
  &=&-\frac{3}{4}Vp_V^2 + V\big[\frac{1}{2}\frac{p_{\Phi}^2}{V^2} + U(\Phi)\big]
  \label{e_hamiltonian}
 \end{eqnarray}
 The above Hamiltonian immediately leads to the Hamiltonian equations, namely (i) the Hamiltonian constraint $\tilde{H} = 0$ gives 
 $H^2 = \frac{1}{3}\big[\frac{1}{2}\dot{\Phi}^2 + U(\Phi)\big]$, where $H$ is 
 the Hubble parameter ($H = \frac{\dot{V}}{3V}$), and (ii) the other Hamiltonian equations $\dot{p}_V = -\frac{\partial\tilde{H}}{\partial V}$ 
 and $\dot{p}_{\Phi} = -\frac{\partial\tilde{H}}{\partial \Phi}$ yield 
 $\big(2\dot{H} + 3H^2\big) + \big(\frac{1}{2}\dot{\Phi}^2 - U(\Phi)\big) = 0$ and $\ddot{\Phi} + 3H\dot{\Phi} + U'(\Phi) = 0$, respectively. 
 The first two equations correspond to the Friedmann equations of Einstein's gravity in the presence of a scalar field, while the last one is 
 the conservation equation of the scalar field. 
 
 \subsection{Improvement coming from holonomy corrections}
Holonomy corrections can be introduced with the replacement of the generalized momenta $p_V$ by 
 $p_V \rightarrow \bigg[-\frac{2}{k\gamma}\sin{\bigg(-\frac{k\gamma}{2}p_V\bigg)}\bigg]$ in the Hamiltonian expression 
 obtained in Eq.~(\ref{e_hamiltonian}). Thereby, the Hamiltonian in  holonomy corrected Einstein's gravity becomes
 \begin{eqnarray}
  \tilde{H} = -\frac{3}{4}V \bigg[-\frac{2}{k\gamma}\sin{\bigg(-\frac{k\gamma}{2}p_V\bigg)}\bigg]^2 
  + V\bigg[\frac{1}{2}\frac{p_{\Phi}^2}{V^2} + U(\Phi)\bigg].
  \label{eh_hamiltonian}
 \end{eqnarray}
 Note that the Hamiltonians for Einstein's gravity  with and without the bholonomy improvement match with each other in the limit $k\gamma \rightarrow 0$. Using Eq.~(\ref{eh_hamiltonian}), we get the corresponding Hamiltonian equations as follows 
 \begin{eqnarray}
  \dot{V}&=&\frac{\partial\tilde{H}}{\partial p_V}\nonumber\\
  \Rightarrow -\frac{2}{k\gamma}\sin{\bigg(-\frac{k\gamma}{2}p_V\bigg)}&=&-\frac{2\dot{V}}{3V\cos{\bigg(-\frac{k\gamma}{2}p_V\bigg)}}~,
  \label{eh1}
 \end{eqnarray}
 \begin{eqnarray}
  \dot{\Phi}&=&\frac{\partial\tilde{H}}{\partial p_{\Phi}}\nonumber\\
  \Rightarrow p_{\Phi}&=&\dot{\Phi}V~,
  \label{eh2}
 \end{eqnarray} 
 \begin{eqnarray}
  \tilde{H}&=&0\nonumber\\
  \Rightarrow H^2&=&\frac{1}{3}\bigg[\frac{1}{2}\dot{\Phi}^2 + U(\Phi)\bigg]\cos^2{\bigg(-\frac{k\gamma}{2}p_V\bigg)}~,
  \label{eh3}
 \end{eqnarray} 
 \begin{eqnarray}
  \dot{p}_V&=&-\frac{\partial\tilde{H}}{\partial V}\nonumber\\
  \Rightarrow 0&=&\frac{2\dot{H}}{\cos^2{\bigg(-\frac{k\gamma}{2}p_V\bigg)} - k\gamma H \tan{\bigg(-\frac{k\gamma}{2}p_V\bigg)}} 
  + \frac{3H^2}{\cos^2{\bigg(-\frac{k\gamma}{2}p_V\bigg)}} + \frac{1}{2}\dot{\Phi}^2 - U(\Phi)~,
  \label{eh4}
 \end{eqnarray}
 and 
 \begin{eqnarray}
  \dot{p}_{\Phi}&=&-\frac{\partial\tilde{H}}{\partial\Phi}\nonumber\\
  \Rightarrow 0&=&\ddot{\Phi} + 3H\dot{\Phi} + U'(\Phi),
  \label{eh5}
 \end{eqnarray}
 where we have use Eq.~(\ref{eh1}), to get
 $\dot{p}_V = \bigg[-\frac{2\dot{H}}{\cos^2{\bigg(-\frac{k\gamma}{2}p_V\bigg)} - k\gamma H \tan{\bigg(-\frac{k\gamma}{2}p_V\bigg)}}\bigg]$. 
 Eqs.~(\ref{eh3}) and (\ref{eh4}) are the Friedmann equations for holonomy improved Einstein's gravity, while Eq.~(\ref{eh5}) corresponds to the dynamics 
 of the scalar field with potential $U(\Phi)$. It should be noticed that the above Friedmann equations are indeed modified due to the holonomy corrections, 
as compared to the case without holonomy corrections. 
 Moreover, the above equations of motion have a term containing the generalized momenta $p_V$. 
 However, this term, $\sin{\big(-\frac{k\gamma}{2}p_V\big)}$, can be replaced with the matter field 
 energy density, by using Eqs.~(\ref{eh1}) and (\ref{eh3}), which immediately lead to 
 $\sin^2{\big(-\frac{k\gamma}{2}p_V\big)} = \frac{\rho}{\rho_c}$, where $\rho = \big[\frac{1}{2}\dot{\Phi}^2 + U(\Phi)\big]$ 
 represents the energy density of the scalar field, and $\rho_c = \frac{3}{k^2\gamma^2}$ is known as the critical energy density. 
 With this expression of $p_V$, the gravitational equations can be written as
 \begin{eqnarray}
  H^2 = \frac{\rho}{3}\bigg[1 - \frac{\rho}{\rho_c}\bigg]
  \label{eh_final1}
 \end{eqnarray}
 and
\begin{eqnarray}
 \dot{H} = -\frac{1}{2}\big(\rho + p\big)\bigg[1 - \frac{2\rho}{\rho_c}\bigg],
 \label{eh_final2}
\end{eqnarray}
with $p = \frac{1}{2}\dot{\Phi}^2 - U(\Phi)$ is the pressure of the scalar field. It is evident that for $k\gamma \rightarrow 0$ (or equivalently 
$\rho_c \rightarrow \infty$), Eqs.~(\ref{eh_final1}) and (\ref{eh_final2}) converge to the usual Friedmann equations 
for usual Einstein's gravity without the holonomy corrections. However it is expected, as for $k\gamma \rightarrow 0$, that
the Hamiltonians with and without holonomy corrections match with each other (as mentioned earlier). 
It is clear, thereby, that the introduction of holonomy corrections 
modify the expressions of $H^2$ and $\dot{H}$ by the terms $\bigg[1 - \frac{\rho}{\rho_c}\bigg]$ and $\bigg[1 - \frac{2\rho}{\rho_c}\bigg]$, respectively, in comparison with the usual Friedmann equations. Thus, in the holonomy corrected scenario, the squared Hubble parameter is proportional to $\rho$ as well as 
$\rho^2$ (apart from some coefficients), which is not the case in the usual FLRW cosmology where $H^2$ is proportional to the linear power of $\rho$, only.

\section{F(R) gravity with a scalar field}\label{sec_F(R)}
This section is devoted to the calculation of the Hamiltonian and its corresponding equations for the F(R) model (along with a scalar field) without 
and with the holonomy improvement (for a general review of $F(R)$ gravity, see e.g., \cite{Capozziello:2011et,Nojiri:2010wj,Nojiri:2017ncd}). 
\subsection{Previous to holonomy corrections}
The Lagrangian of the F(R) model with a scalar field in the FLRW spacetime geometry is given by
\begin{eqnarray}
 L = \frac{1}{2}VF(R) + V\big(\frac{1}{2}\dot{\Phi}^2 - U(\Phi)\big)
 \label{FL1}
\end{eqnarray}
where $V = a^3$ is the volume and $R$ is the scalar curvature. It may be mentioned that a F(R) model can be equivalently mapped to a scalar-tensor theory by using a conformal transformation of the metric, where the scalar field potential of the ST theory depends on the form of F(R). 
However, here we are interested in obtaining the Hamiltonian in the Jordan frame F(R) model and thus we stick to the Lagrangian shown in Eq.~(\ref{FL1}) 
(see \cite{haro1}, for the Hamiltonian formalism of the F(R) model in the Einstein frame). Similar to Einstein's gravity, here we introduce the Lagrange multiplier $\mu$ in the above Lagrangian $L$, to get
\begin{eqnarray}
 L_1 = \frac{1}{2}VF(R) + V\big(\frac{1}{2}\dot{\Phi}^2 - U(\Phi)\big) + \frac{1}{2}\mu\bigg[\frac{2\ddot{V}}{V} - \frac{2\dot{V}^2}{3V^2} - R\bigg]
 \label{FL2}
\end{eqnarray}
Variation of $R$ gives $\mu = \frac{1}{2}VF'(R)$, where the prime denotes differentiation with respect to $R$. Further, to remove the second derivative 
of $V$ we subtract a total derivative term $\frac{d}{dt}\big(F'(R)\dot{V}$ from $L_1$, as follows
\begin{eqnarray}
 \tilde{L}&=&L_1 - \frac{d}{dt}\big(F'(R)\dot{V}\big)\nonumber\\
 &=&\frac{1}{2}VF(R) - \frac{1}{2}VF'(R)\big(\frac{2\dot{V}^2}{3V^2} + R\big) - F''(R)\dot{R}\dot{V} + V\big(\frac{1}{2}\dot{\Phi}^2 - U(\Phi)\big).
 \label{FL3}
\end{eqnarray}
Note that the final Lagrangian $\tilde{L}$ depends on the variables $(V,R,\Phi)$ and their first derivatives. The corresponding conjugate momenta can 
be expressed as
\begin{eqnarray}
 p_V&=&\frac{\partial\tilde{L}}{\partial\dot{V}} = -\frac{2}{3}F'(R)\frac{\dot{V}}{V} - F''(R)\dot{R}\nonumber\\
p_R&=&\frac{\partial\tilde{L}}{\partial\dot{R}} = - F''(R)\dot{V}\nonumber\\
p_{\Phi}&=&\frac{\partial\tilde{L}}{\partial\dot{\Phi}} = V\dot{\Phi},\nonumber\\
\nonumber
\end{eqnarray}
respectively. These expressions of canonical momenta along with the Lagrangian $\tilde{L}$ yield the Hamiltonian for ordinary F(R) gravity (without holonomy 
corrections) as
\begin{eqnarray}
 \tilde{H}&=&p_v\dot{V} + p_R\dot{R} + p_{\Phi}\dot{\Phi} - \tilde{L}\nonumber\\
 &=&\frac{V}{2}\big[RF'(R) - F\big] - \frac{p_Vp_R}{F''(R)} + \frac{F'(R)}{3V(F'')^2}p_R^2 + V\big[\frac{1}{2}\frac{p_{\Phi}^2}{V^2} + U(\Phi)\big]
 \label{F_hamiltonian}
\end{eqnarray}
The Hamiltonian constraint $\tilde{H} = 0$ leads to the well known Friedmann equation in the F(R) model, namely
\begin{eqnarray}
 \tilde{H}&=&0\nonumber\\
 \Rightarrow -\frac{F}{2}&+&F'(R)\big[\frac{1}{2}R - 3H^2\big] - 3H\frac{dF'}{dt} + \rho = 0.
 \label{F1}
\end{eqnarray}
The other Friedmann equation and the scalar field equation are obtained from the other Hamiltonian equations, as follows 
\begin{eqnarray}
 \dot{p}_V&=&-\frac{\partial\tilde{H}}{\partial V}\nonumber\\
 \Rightarrow -\frac{1}{2}\big[RF'(R) - F\big]&+&\big(2\dot{H} + 3H^2\big)F'(R) + 2H\frac{dF'}{dt} + \frac{d^2F'}{dt^2} 
 + \big(\frac{1}{2}\dot{\Phi}^2 - U(\Phi)\big) = 0
 \label{F2}
\end{eqnarray}
and
\begin{eqnarray}
 \dot{p}_{\Phi}&=&-\frac{\partial\tilde{H}}{\partial\Phi}\nonumber\\
 \Rightarrow \ddot{\Phi}&+&3H\dot{\Phi} + U'(\Phi) = 0,
 \label{F3}
\end{eqnarray}
respectively. In the following subsection, we will explore the Hamiltonian formalism of F(R) gravity in the presence of holonomy corrections, and examine how such corrections modify the Friedmann equations, in comparison with the usual F(R) case (without holonomy corrections).

\subsection{Holonomy improvement}
Performing the holonomy corrections 
$p_V \rightarrow \bigg[-\frac{2}{k\gamma}\sin{\bigg(-\frac{k\gamma}{2}p_V\bigg)}\bigg]$ in Eq.~(\ref{F_hamiltonian}), the Hamiltonian of F(R) gravity  takes the following form,
\begin{eqnarray}
 \tilde{H} = 
 \frac{V}{2}\big[RF'(R) - F\big] - \frac{p_R}{F''(R)}\big[-\frac{2}{k\gamma}\sin{\bigg(-\frac{k\gamma}{2}p_V\bigg)}\big] 
 + \frac{F'(R)}{3V(F'')^2}p_R^2 + V\big[\frac{1}{2}\frac{p_{\Phi}^2}{V^2} + U(\Phi)\big]
 \label{Fh_hamiltonian}
\end{eqnarray}
The above expression for $\tilde{H}$ immediately leads to the corresponding Hamiltonian equations, as
\begin{eqnarray}
 \dot{V}&=&\frac{\partial\tilde{H}}{\partial p_V}\nonumber\\
 \Rightarrow p_R&=&-F''(R)\dot{V}/\cos{b}~,
\label{Fh2}
\end{eqnarray}
\begin{eqnarray}
 \dot{\Phi} = \frac{\partial\tilde{H}}{\partial p_{\Phi}} = \frac{p_{\Phi}}{V}~,
 \label{Fh3}
\end{eqnarray}
\begin{eqnarray}
 \tilde{H}&=&0\nonumber\\
 \Rightarrow -\frac{F(R)}{2}&+&F'(R)\big[\frac{1}{2}R - \frac{3H^2}{\cos^2{b}}\big] - 3H\frac{dF'}{dt}\frac{1}{\cos{b}} + \big(\frac{1}{2}\dot{\Phi}^2 
 + U(\Phi)\big) = 0~,
 \label{Fh4}
\end{eqnarray}
\begin{eqnarray}
 \dot{p}_V&=&-\frac{\partial\tilde{H}}{\partial V}\nonumber\\
 \Rightarrow 0&=&\frac{\frac{d^2F'}{dt^2}\cos{b}+2\dot{H}F'+2H\frac{dF'}{dt}}{\cos^2{b}-k\gamma HF'(R)\tan{b}} - \frac{1}{2}\big[RF' - F\big] 
 + \big(\frac{1}{2}\dot{\Phi}^2 - U(\Phi)\big) + \frac{3H^2F'}{\cos^2{b}}~,
 \label{Fh5}
\end{eqnarray}
and
\begin{eqnarray}
 \dot{p}_{\Phi}&=&-\frac{\partial\tilde{H}}{\partial\Phi}\nonumber\\
 \Rightarrow \ddot{\Phi}&+&3H\dot{\Phi} + U'(\Phi) = 0~,
 \label{Fh6}
\end{eqnarray}
where $b = -\frac{k\gamma}{2}p_V$ and, moreover, we use Eqn.~(\ref{Fh2}) to get 
 $\dot{p}_V = \frac{-\frac{d^2F'}{dt^2}\cos{b} - 2\dot{H}F' - 2H\frac{dF'}{dt}}{\cos^2{b} - k\gamma HF'(R)\tan{b}}$. The first three  equations 
 yield the  generalized momenta, while Eqns.~(\ref{Fh4}), (\ref{Fh5}) are the Friedmann equations of F(R) gravity in the holonomy corrected scenario, and 
 Eqn.~(\ref{Fh6}) is the conservation equation of the scalar field. Note that in the limit $k\gamma \rightarrow 0$, all the 
 above expressions match the equations of the F(R) model without holonomy corrections. 
 Similarly to Einstein's gravity, here we also represent the generalized momenta $p_V$ in terms of the scalar field energy density 
($\rho = \frac{1}{2}\dot{\Phi}^2 + U$) and the form of F(R). Following \cite{haro1}, we easily get
\begin{eqnarray}
 \sin^2{b} = \frac{3A}{4\rho_c},
 \nonumber
\end{eqnarray}
where $A = \big(\dot{R}F''\big)^2 + \frac{2}{3}F'\big(RF' - F + 2\rho\big)$ and $\rho_c = \frac{3}{k^2\gamma^2}$. Using this expression, 
Eqns.~(\ref{Fh4}) and (\ref{Fh5}) take the form 
\begin{eqnarray}
 -\frac{F(R)}{2} + F'(R)\bigg[\frac{1}{2}R - \frac{3H^2}{1-\frac{3A}{4\rho_c}}\bigg] - 3H\frac{dF'}{dt}\frac{1}{\sqrt{1-\frac{3A}{4\rho_c}}} 
 + \big(\frac{1}{2}\dot{\Phi}^2 + U(\Phi)\big) = 0
 \label{Fh_final1}
\end{eqnarray}
and
\begin{eqnarray}
 \frac{\frac{d^2F'}{dt^2}\sqrt{1-3A/4\rho_c} + 2\dot{H}F' + 2H\frac{dF'}{dt}}{(1-3A/4\rho_c) - k\gamma HF'(R)\frac{\sqrt{3A/4\rho_c}}{\sqrt{1-3A/4\rho_c}}} 
 - \frac{1}{2}\big[RF' - F\big] + \big(\frac{1}{2}\dot{\Phi}^2 - U(\Phi)\big) + \frac{3H^2F'}{1 - 3A/4\rho_c} = 0,
 \label{Fh_final2}
\end{eqnarray}
respectively. It is clear that the holonomy corrections vanish in the limit $k\gamma \rightarrow 0$, as expected.
 
In a matter or qausi-matter bouncing universe, the holonomy corrections may have significant imprints near the bouncing point, 
however in the deep contracting era, i.e., 
at large negative time when the spacetime perturbations are generated, the scalar curvature is large as compared to that of the bouncing era, and thus 
the holonomy corrections may be safely disregarded in the matter dominated epoch. 
In most of the previous models, the energy conditions have to be violated, in order to get a non-singular bounce. However in the present paper, we 
are mainly interested in whether the holonomy corrections may restore the energy conditions in the background of matter or quasi-matter bounce scenario. To this purpose, we consider the Lagrange multiplier F(R) gravity model as it yields a viable phenomenology in a matter (or quasi-matter) bouncing universe, 
unlike the case of pure F(R) gravity, as explored in our earlier paper \cite{tp_bouncing1}. Keeping this in mind, in the next section we shall determine the Hamiltonian and the corresponding equations of the  F(R) model with Lagrange multiplier, without and with holonomy corrections, in order to explicitly characterize the modifications generated by the holonomy improvement.

\section{F(R) gravity with Lagrange multiplier}\label{sec_LMC}

\subsection{No holonomy correction}
Let us briefly recall the formalism of F(R) gravity with Lagrange multiplier, developed in Ref.~\cite{Nojiri:2017ygt}. The action
of the model is,
\begin{align}
S = \frac{1}{2\kappa^2} \int d^4x \sqrt{-g}\left[F(R) 
+ \lambda\left(\partial_{\mu}\Phi\partial^{\mu}\Phi 
+ G(R)\right)\right]\, ,
\label{action1}
\end{align}
where $\kappa^2 = \frac{1}{M^2}$, with $M$ the four dimensional
Planck mass $\sim 10^{19}$ GeV. Here, $F(R)$ and $G(R)$ are two
differentiable functions of the Ricci scalar $R$, $\Phi$ is a scalar
field with a self coupling kinetic term and the coupling is
determined by the function $\lambda$, known as the Lagrange
multiplier, in the action (\ref{action1}). It was shown in
\cite{Nojiri:2017ygt}, that this variant of $F(R)$ gravity
with a Lagrange multiplier term is free of ghosts. However, the Lagrangian contains a higher derivative of $V = a^3$ and thus the suitable form 
of the Lagrangian for determining the Hamiltonian is given by
\begin{eqnarray}
 \tilde{L}&=&\frac{1}{2}V \bigg[F(R) + \lambda\big(G(R) - \dot{\Phi}^2\big)\bigg] 
 + \frac{1}{2}Vf_R\bigg[\frac{2\ddot{V}}{V} - \frac{2\dot{V}^2}{3V^2} - R\bigg] - \frac{d}{dt}\big(f_R\dot{V}\big)\nonumber\\
 &=&\frac{1}{2}Vf - \frac{1}{2}V\lambda\dot{\Phi}^2 - \frac{1}{2}Vf_R\bigg[\frac{2\dot{V}^2}{3V^2}+R\bigg] - f_{RR}\dot{R}\dot{V} - G_R\dot{\lambda}\dot{V}
 \label{ML3}
\end{eqnarray}
with $f(R,\lambda) = F(R) + \lambda G(R)$, and where we denote $f_R = df/dR$ (the same notation for $G(R)$). It should be observed that the above Lagrangian 
depends on the variables $(V,R,\lambda,\Phi)$, thus the canonical momenta turns out to be
\begin{eqnarray}
 p_V&=&\frac{\partial\tilde{L}}{\partial\dot{V}} = -f_{RR}\dot{R} - \frac{2\dot{V}}{3V}f_R - G_R\dot{\lambda}\nonumber\\
 p_R&=&\frac{\partial\tilde{L}}{\partial\dot{R}} = -f_{RR}\dot{V}\nonumber\\
 p_{\lambda}&=&\frac{\partial\tilde{L}}{\partial\dot{\lambda}} = -G_R\dot{V}. \nonumber\\
 p_{\Phi}&=&\frac{\partial\tilde{L}}{\partial\dot{\Phi}} = -V\lambda\dot{\Phi}.
 \nonumber
\end{eqnarray}
Consequently the Hamiltonian is given by,
\begin{eqnarray}
 \tilde{H}&=&p_V\dot{V} + p_R\dot{R} + p_{\lambda}\dot{\lambda} + p_{\Phi}\dot{\Phi} - \tilde{L}\nonumber\\
 &=&\frac{V}{2}\big[Rf_R - f\big] - \frac{1}{2}p_V\bigg(\frac{p_R}{f_{RR}} + \frac{p_{\lambda}}{G_R}\bigg) 
 + \frac{f_R}{12V}\bigg(\frac{p_R}{f_{RR}} + \frac{p_{\lambda}}{G_R}\bigg)^2 - \frac{1}{2}\frac{p_{\Phi}^2}{V\lambda}.
 \label{M_hamiltonian}
\end{eqnarray}
The Hamiltonian constraint $\tilde{H} = 0$ and the other Hamiltonian equation $\dot{p}_V = -\frac{\partial\tilde{H}}{\partial V}$ lead 
to the well-known Friedmann equations for Lagrange multiplier F(R) gravity without the holonomy corrections \cite{Nojiri:2017ygt}, as
\begin{eqnarray}
  -\frac{F}{2} - \frac{\lambda}{2}\big(\dot{\Phi}^2 + G\big) + \big(F_R + \lambda G_R\big)\big(\frac{R}{2}-3H^2\big) 
 - 3H\frac{d}{dt}\big(F_R + \lambda G_R\big) = 0
 \label{M1}
\end{eqnarray}
and
\begin{eqnarray}
 \frac{F}{2} + \frac{\lambda}{2}\big(G - \dot{\Phi}^2\big) + \big(F_R + \lambda G_R\big)\big(2\dot{H}+3H^2-\frac{R}{2}\big) 
 + 2H\frac{d}{dt}\big(F_R + \lambda G_R\big) + \frac{d^2}{dt^2}\big(F_R + \lambda G_R\big) = 0,
 \label{M2}
\end{eqnarray}
respectively. The dynamics corresponding to the fields $\Phi$ and $\lambda$ are obtained from $\dot{p}_{\Phi} = -\frac{\partial\tilde{H}}{\partial\Phi}$ and 
$\dot{p}_{\lambda} = -\frac{\partial\tilde{H}}{\partial\lambda}$, as follows
\begin{eqnarray}
 \dot{p}_{\Phi}&=&-\frac{\partial\tilde{H}}{\partial\Phi}\nonumber\\
 \Rightarrow \lambda&=&\frac{E}{a^3\dot{\Phi}}
 \label{M3}
\end{eqnarray}
and
\begin{eqnarray}
 \dot{p}_{\lambda}&=&-\frac{\partial\tilde{H}}{\partial\lambda}\nonumber\\
 \Rightarrow \dot{\Phi}^2&=&G(R),
 \label{M4}
\end{eqnarray}
respectively, with $E$ being an integration constant. Taking eqn.(\ref{M4}) into account, the Lagrange multiplier can be written as 
$\lambda = \frac{E}{a^3\sqrt{G}}$. As a result, the gravitational equations take the  form
\begin{eqnarray}
 -&\frac{F}{2}& - \frac{E\sqrt{G}}{a^3} + \big(F_R + \frac{EG_R}{a^3\sqrt{G}}\big)\big(\frac{R}{2}-3H^2\big) 
 - 3H\frac{d}{dt}\big(F_R + \frac{EG_R}{a^3\sqrt{G}}\big) = 0\nonumber\\
 &\frac{F}{2}& + \big(F_R + \frac{EG_R}{a^3\sqrt{G}}\big)\big(2\dot{H}+3H^2-\frac{R}{2}\big) 
 + 2H\frac{d}{dt}\big(F_R + \frac{EG_R}{a^3\sqrt{G}}\big) + \frac{d^2}{dt^2}\big(F_R + \frac{EG_R}{a^3\sqrt{G}}\big) = 0.
 \label{M_final}
\end{eqnarray}
Eqn.~(\ref{M_final}) denotes the final Friedmann equations for Lagrange multiplier F(R) gravity in absence of holonomy corrections, which also 
match with \cite{Nojiri:2017ygt}.

\subsection{Holonomy improvement}
Similarly as in previous cases, the holonomy corrections can be introduced by the replacement 
$p_V \rightarrow \bigg[-\frac{2}{k\gamma}\sin{\bigg(-\frac{k\gamma}{2}p_V\bigg)}\bigg]$ in eqn.(\ref{M_hamiltonian}) and thus the modified 
Hamiltonian takes the form
\begin{eqnarray}
 \tilde{H} = \frac{V}{2}\big[Rf_R - f\big] 
 - \frac{1}{2}\bigg(\frac{p_R}{f_{RR}} + \frac{p_{\lambda}}{G_R}\bigg)\bigg[-\frac{2}{k\gamma}\sin{b}\bigg] 
 + \frac{f_R}{12V}\bigg(\frac{p_R}{f_{RR}} + \frac{p_{\lambda}}{G_R}\bigg)^2 - \frac{1}{2}\frac{p_{\Phi}^2}{V\lambda}
 \label{Mh_hamiltonian}
\end{eqnarray}
with $b = -\frac{k\gamma}{2}p_V$. Eqn.~(\ref{Mh_hamiltonian}) clearly evidences that $\tilde{H}$ depends on the variables 
$(V,R,\lambda,\Phi)$ and their conjugate momenta. The dynamics corresponding to $(V,R,\lambda,\Phi)$ can be obtained as
\begin{eqnarray}
 \dot{V}&=&\frac{\partial\tilde{H}}{\partial p_V} = -\frac{1}{2}\bigg(\frac{p_R}{f_{RR}} + \frac{p_{\lambda}}{G_R}\bigg)\cos{b}\nonumber\\
 \dot{R}&=&\frac{\partial\tilde{H}}{\partial p_R} = 
 \frac{1}{k\gamma f_R}\sin{b} + \frac{f_R}{6Vf_{RR}}\bigg(\frac{p_R}{f_{RR}} + \frac{p_{\lambda}}{G_R}\bigg)\nonumber\\
\dot{\lambda}&=&\frac{\partial\tilde{H}}{\partial p_{\lambda}} = 
\frac{1}{k\gamma G_R}\sin{b} + \frac{f_R}{6Vf_{RR}}\bigg(\frac{p_R}{G_R} + \frac{p_{\lambda}}{G_R}\bigg)\nonumber\\
\dot{\Phi}&=&\frac{\partial\tilde{H}}{\partial p_{\Phi}} = -\frac{p_{\Phi}}{v\lambda}.
\label{Mh1}
\end{eqnarray}
Moreover, the Hamiltonian constraint $\tilde{H} = 0$ and the other Hamiltonian equation $\dot{p}_V = -\frac{\partial\tilde{H}}{\partial V}$ yield the 
Friedmann equations for the Lagrange multiplier F(R) gravity model in presence of holonomy corrections, as
\begin{eqnarray}
 \tilde{H}&=&0\nonumber\\
 \Rightarrow -\frac{F}{2} - \frac{\lambda}{2}\big(\dot{\Phi}^2 + G\big)&+&\big(F_R + \lambda G_R\big)\big(\frac{R}{2} - 3H^2/\cos^2{b}\big) 
 - \bigg(\frac{3H}{\cos{b}}\bigg)\frac{d}{dt}\big(F_R + \lambda G_R\big) = 0
 \label{Mh2}
\end{eqnarray}
and
\begin{eqnarray}
 \dot{p}_V&=&-\frac{\partial\tilde{H}}{\partial V}\nonumber\\
 \Rightarrow \frac{F}{2} + \frac{\lambda}{2}\big(G - \dot{\Phi}^2\big)&+&\big(F_R + \lambda G_R\big)\big(\frac{3H^2}{\cos^2{b}}-\frac{R}{2}\big) 
 + \frac{\frac{d^2f_R}{dt^2}\cos{b}+2\dot{H}f_R+2H\frac{df_R}{dt}}{\cos^2{b}-k\gamma Hf_R\tan{b}} = 0,
 \label{Mh3}
\end{eqnarray}
respectively, where we have used Eq.~(\ref{Mh1}). The other two equations of motion, $\dot{p}_{\Phi} = -\frac{\partial\tilde{H}}{\partial\Phi}$ and 
$\dot{p}_{\lambda} = -\frac{\partial\tilde{H}}{\partial\lambda}$, lead to $\lambda = \frac{E}{a^3\dot{\Phi}}$ (with $E$ an integration constant) 
and $\dot{\Phi}^2 = G(R)$, respectively. To complete the equations of motion, we need to represent $\sin{b}$ in terms of the scalar curvature and the scalar field energy density. Using the above equations, we get (see Appendix-I)
\begin{eqnarray}
 \sin^2{b} = \frac{3A}{4\rho_c},
 \label{appendix}
\end{eqnarray}
with $A = 4\big(\dot{R}f_{RR}\big)^2 + \frac{2}{3}f_R\big(Rf_R - f - \lambda\dot{\Phi}^2\big)$ and $\rho_c = \frac{3}{k^2\gamma^2}$. This expression of 
$\sin{b}$ leads to the final form of the gravitational equations, as folllows
\begin{eqnarray}
 -\frac{F}{2} - \frac{E\sqrt{G}}{a^3} + \bigg(F_R + \frac{EG_R}{a^3\sqrt{G}}\bigg)\bigg(\frac{R}{2} - \frac{3H^2}{1-\frac{3A}{4\rho_c}}\bigg) 
 - \bigg(\frac{3H}{\sqrt{1-3A/4\rho_c}}\bigg)\frac{d}{dt}\big(F_R + \frac{EG_R}{a^3\sqrt{G}}\big) = 0
 \label{Mh_final1}
\end{eqnarray}
and
\begin{eqnarray}
 \frac{F}{2} + \bigg(F_R + \frac{EG_R}{a^3\sqrt{G}}\bigg)\bigg(\frac{3H^2}{1-3A/4\rho_c} - \frac{R}{2}\bigg) 
 + \frac{\frac{d^2f_R}{dt^2}\sqrt{1-3A/4\rho_c} + 2\dot{H}f_R + 2H\frac{df_R}{dt}}
 {(1-3A/4\rho_c) - k\gamma Hf_R\frac{\sqrt{3A/4\rho_c}}{\sqrt{1-3A/4\rho_c}}} = 0,
 \label{Mh_final2}
\end{eqnarray}
respectively. Comparing Eqs.~(\ref{Mh_final1}) and (\ref{Mh_final2}) with Eq.~(\ref{M_final}), we immediately identify the modification to the Friedmann 
equations generated by the holonomy improvement. As mentioned earlier, such modifications become significant near the bouncing era in a bouncing universe. 
We investigate whether the holonomy modifications can restore the energy conditions in the backdrop of a matter or quasi-matter bouncing scenario. 
However, before the investigation of the energy condition, it is crucial to determine the observable parameters (like the spectral index and tensor to scalar ratio) and directly confront them with the values coming from the most recent Planck observations, which is the subject of next section.

\section{Realization of the  bouncing cosmology}\label{sec_reconstruction}
In the present section we will determine the observable quantities for the bouncing universe described by the scale factor 
\begin{eqnarray}
 a(t) = \left(a_0t^2 + 1 \right)^n,
 \label{scale factor}
\end{eqnarray}
where $a_0$ and $n$ are the free parameters of the model,  with $a_0$
having mass dimension [+2], while $n$ is dimensionless. It must be mentioned that, for
$n=1/3$, the scale factor describes a matter bounce scenario. The
Universe's evolution in a general bouncing cosmology consists of
two characteristic eras: an era of contraction and one of expansion. It is
obvious that the above scale factor describes a contracting era
for the Universe, when $t\to -\infty$; then the Universe reaches a
bouncing point, at $t=0$, at which the Universe has a minimal size,
and afterwards the Universe starts to expand again, for cosmic times
$t>0$. Hence, the Universe in this scenario never develops a
crushing type Big Bang singularity.

For the purpose of determining the observable quantities, we will consider the low curvature limit of the theory. So, before proceeding further, 
let us comment on the viability of this approximation. We do it in the context of matter bounce cosmology, which is obtained by taking $n=1/3$ in 
Eq.~(\ref{scale factor}), the primordial perturbations of the comoving
curvature, which originate from quantum vacuum fluctuations, were
at subhorizon scales during the contracting era in the
low-curvature regime, that is, their wavelength was much smaller
than the comoving Hubble radius, which is defined by $r_h =
\frac{1}{aH}$. In the matter bounce evolution, the Hubble horizon
radius decreases in size, and this causes the perturbation modes
to exit from the horizon eventually, with this exit occurring when
the contracting Hubble horizon becomes equal to the wavelength of
these primordial modes. However, in the present context, we
consider a larger class of bouncing models of the form $a(t) =
(a_0t^2 + 1)^n$, in the presence of a generalized Lagrange
multiplier $F(R)$ gravity. Thus, it will be important to check what are the
possible values of $n$ which make the low-curvature limit, that is, $R/a_0 \ll 1$ a viable
approximation in calculating the power spectrum for the bouncing
model $a(t) = (a_0t^2 + 1)^n$. This expression of the scale factor immediately leads to the comoving Hubble radius
\begin{align}
r_h = \frac{(1 + a_0t^2)^{1-n}}{2a_0nt} \, .
\label{viability1}
\end{align}
Thereby $r_h$ diverges at $t \simeq 0$, as expected because the
Hubble rate goes to zero at the bouncing point. Furthermore,
the asymptotic behavior of $r_h$ is given by $r_h \sim t^{1-2n}$,
thus $r_h\left(|t|\rightarrow \infty\right)$ diverges for $n < 1/2$,
otherwise $r_h$ goes to zero asymptotically. Hence, for $n < 1/2$,
the comoving Hubble radius decreases initially in the contracting
era and then diverges near the bouncing point, unlike in the case
$n > 1/2$, where the Hubble radius increases from the past infinite
and gradually diverges at $t = 0$. As a result, the possible range
of $n$ which leads the perturbation modes to exit the horizon at
large negative time and make the low-curvature limit a viable
approximation in calculating the power spectrum, is given by $0 <
n < 1/2$. Moreover we will show in the later sections that this range of $n$ makes the observable quantities compatible 
with the Planck constraints and, thus, the ``low curvature limit'' comes as a viable approximation to calculate the power spectra of scalar and tensor 
perturbations. Further, in such low curvature regime (i.e in the deep contracting era), the holonomy corrections may be safely disregarded, which in turn makes the Friedmann equations (see Eqs.~(\ref{Mh_final1}) and (\ref{Mh_final2}), as follows
\begin{eqnarray}
 -\frac{F}{2} - \frac{E\sqrt{G}}{a^3} + \big(F_R + \frac{EG_R}{a^3\sqrt{G}}\big)\big(\frac{R}{2}-3H^2\big) 
 - 3H\frac{d}{dt}\big(F_R + \frac{EG_R}{a^3\sqrt{G}}\big) = 0
 \label{Mh_final11}
 \end{eqnarray}
 and
 \begin{eqnarray}
 \frac{F}{2} + \big(F_R + \frac{EG_R}{a^3\sqrt{G}}\big)\big(2\dot{H}+3H^2-\frac{R}{2}\big) 
 + 2H\frac{d}{dt}\big(F_R + \frac{EG_R}{a^3\sqrt{G}}\big) + \frac{d^2}{dt^2}\big(F_R + \frac{EG_R}{a^3\sqrt{G}}\big) = 0,
 \label{Mh_final22}
\end{eqnarray}
respectively. Thereby, the Lagrange multiplier F(R) gravity model with/without the holonomy corrections behave similarly in respect to 
the observable quantities, in the backdrop of a matter or quasi-matter bounce scenario. With these equations, we shall investigate which functional
forms of $F(R)$ and $G(R)$ can realize a bouncing Universe
cosmological scenario, with the scale factor shown in Eq.~(\ref{scale factor}), which leads to the following Hubble rate and its first derivative
\begin{align}
H(t) = \frac{2nt}{t^2 + 1/a_0 }\, , \quad 
\dot{H}(t) = -2n\frac{t^2 - 1/a_0}{\left(t^2 +1/a_0\right)^2}\, .
\end{align}
With the help of the above expressions, the Ricci scalar is found
to be
\begin{align}
R(t)=12H^2 + 6\dot{H} 
=12n \left[\frac{(4n-1)t^2 + 1/a_0}{\left(t^2 + 1/a_0\right)^2}\right] \, .
\label{ricci scalar}
\end{align}
Using Eq.~(\ref{ricci scalar}), one can determine the cosmic time
as a function of the Ricci scalar, that is, the function $t = t(R)$. As
a result, the Hubble rate and its first derivative can be
expressed in terms of $R$ (however, this statement holds for all
analytic functions of $t$), and also the differential operator
$\frac{d}{dt}$ can be written as $\frac{d}{dt} = \dot{R}(R)
\frac{d}{dR}$. By plugging the resulting expressions into 
Eqs.~(\ref{Mh_final11}) and (\ref{Mh_final22}), we obtain differential
equations which determine the functional forms of $F(R)$, $G(R)$
fully in terms of $R$, and by solving those differential
equations, the forms of $F(R)$ and $G(R)$ can be found.

During the low-curvature regime ($\frac{R}{a_0} \ll 1$, or for a large negative time),
$R(t)$ can be written as $R(t) \sim \frac{12n(4n-1)}{t^2}$ from
Eq.~(\ref{ricci scalar}). This helps to express the scale factor,
the Hubble rate, its first derivative, and the differential
operators $d/dt$, $d^2/dt^2$, in terms of  the Ricci scalar $R$, as
follows
\begin{align}
a(R) = \frac{\left[12na_0^n(4n-1)\right]^n}{R^n} \, , \quad 
H(R) = 2n\sqrt{\frac{R}{12n(4n-1)}} \, , \quad 
\dot{H}(R)= -2n\sqrt{\frac{R}{12n(4n-1)}}\, ,
\end{align}
and
\begin{align}
 \frac{d}{dt}=-24n(4n-1)\left[\frac{R}{12n(4n-1)}\right]^{3/2} \frac{d}{dR}
\, , \quad 
\frac{d^2}{dt^2}= \frac{1}{3n(4n-1)} \left[R^3\frac{d^2}{dR^2} 
+ \frac{3}{2}R^2\frac{d}{dR}\right]\, ,
\label{low curvature quantites}
\end{align}
respectively. By plugging back these expressions into 
Eqs.~(\ref{Mh_final11}) and (\ref{Mh_final22}), and by introducing 
$J(R) = F(R) + \frac{2E\sqrt{G(R)}}{a^3(R)}$, we get the following differential
equations,
\begin{align}
\frac{2}{(4n-1)}R^2 \frac{d^2J}{dR^2} - \frac{(1-2n)}{(4n-1)}R\frac{dJ}{dR} - J(R) 
= 0\, ,
\label{low curvature eqn1}
\end{align}
and
\begin{align}
F(R) = \frac{(6n-1)}{3n(4n-1)}R\frac{dJ}{dR} - \frac{(3-4n)}{3n(4n-1)}R^2\frac{d^2J}{dR^2} 
 - \frac{2}{3n(4n-1)}R^3\frac{d^3J}{dR^3}\, .
\label{low curvature eqn2}
\end{align}
Eq.~(\ref{low curvature eqn1}) has the following solution
\begin{align}
J(R) = A R^{\rho} + B R^{\delta} \, , 
\label{low curvature sol1}
\end{align}
where $\rho = \frac{1}{4}\left[3 - 2n - \sqrt{1 + 4n(5+n)}\right]$
and $\delta = \frac{1}{4}\left[3 - 2n + \sqrt{1 + 4n(5+n)}\right]$,
and also $A$ and $B$ are integration constants having mass dimensions
$[2-2\rho]$ and $[2-2\delta]$, respectively. This solution of
$J(R)$ along with Eq.~(\ref{low curvature eqn2}), lead to the
following functional form of $F(R)$
\begin{align}
F(R)=&A \left[\frac{(6n-1)}{3n(4n-1)}\rho - \frac{(3-4n)}{3n(4n-1)}\rho(\rho-1) 
 - \frac{2}{3n(4n-1)}\rho(\rho-1)(\rho-2)\right]R^{\rho}\nonumber\\
&+B \left[\frac{(6n-1)}{3n(4n-1)}\delta - \frac{(3-4n)}{3n(4n-1)}\delta(\delta-1) 
 - \frac{2}{3n(4n-1)}\delta(\delta-1)(\delta-2)\right]R^{\delta}\nonumber\\
=&C R^{\rho} + D R^{\delta}\, ,
\label{low curvature sol2}
\end{align}
where $C$ and $D$ are the corresponding coefficients of $R^{\rho}$
and $R^{\delta}$, respectively. With these solutions, the effective
$f(R)$ can be written as,
\begin{align}
f(R)= F(R) + \lambda G(R) 
= \frac{1}{2}\left[J(R) + F(R)\right] 
= \frac{1}{2}(A+C) R^{\rho} + \frac{1}{2}(B+D) R^{\delta}
\label{low curvature sol3}
\end{align}
where we have used the solution of $\lambda(t) = \frac{E}{a^3\sqrt{G}}$.

Eqs.~(\ref{low curvature sol1}), (\ref{low curvature sol2}), 
and (\ref{low curvature sol3}) constitute the main results of the present
section. It is interesting to note that due to well-known equivalence of F(R) gravity with generalized fluid models (see \cite{Nojiri:2006ri}), 
the same scenario may be induced by a specific generalized fluid. In the next section we address concretely the
cosmological perturbations issue and we shall confront the theory
with the observational data.


\section{Cosmological perturbation: Observable quantities}\label{sec_perturbation}

In this section we shall study the first order metric
perturbations of the theory at hand, following 
Refs.~\cite{hwang1,hwang2,hwang3}, where the scalar and tensor
perturbations are calculated for various variants of higher
curvature gravity models. Scalar, vector and tensor perturbations
are decoupled, as in general relativity, so that we can focus our
attention to tensor and scalar perturbations
separately.

\subsection{Scalar perturbations}

A scalar perturbation of the FRW background can be written as
\begin{align}
ds^2 = -(1 + 2\Psi)dt^2 + a(t)^2(1 - 2\Psi)\delta_{ij}dx^{i}dx^{j}\, ,
\label{sp1}
\end{align}
where $\Psi(t,\vec{x})$ denotes the perturbation. In
principle, perturbations should always be expressed in
terms of gauge invariant quantities, namely in our case the comoving
curvature perturbation, defined as $\Re = \Psi - aHq$, where
$v(t,\vec{x})$ is the velocity perturbation. However, we shall
work in the comoving gauge condition, where the velocity
perturbation is taken to be zero; with such gauge fixing $\Re =
\Psi$. Thereby, we can work with the perturbed variable
$\Psi(t,\vec{x})$. The perturbed action up to $\Psi^2$ order is
\cite{hwang1},
\begin{align}
\delta S_{\psi} = \int dt d^3\vec{x} a(t) z(t)^2\left[\dot{\Psi}^2 
 - \frac{1}{a^2}\left(\partial_i\Psi\right)^2\right]\, ,
\label{sp2}
\end{align}
where $z(t)$ has the following expression
\begin{align}
z(t) = \frac{a(t)}{\left(H(t) 
+ \frac{1}{2f'(R)}\frac{df'(R)}{dt}\right)} \sqrt{\frac{2E\sqrt{G}}{a^3} 
+ \frac{3}{2f'(R)}\left(\frac{df'(R)}{dt}\right)^2}\, .
\label{sp3}
\end{align}
It is plain, from Eq.~(\ref{sp2}), that $c_s^2 = 1$, which
guarantees the absence of ghost modes quantified in terms of
superluminal propagating modes of the model. Recall that the low curvature approximation stands as a viable check in calculating the observable quantities. 
Thus, in the low-curvature limit, we determine various terms present in
the expression of $z(t)$ (see Eq.~(\ref{sp3})), as
\begin{align}
\frac{a(t)}{\left(H(t) + \frac{1}{2f'(R)}\frac{df'(R)}{dt}\right)} 
= \frac{a_0^n\left(12n(4n-1)\right)^{n+1/2}}{R^{n+1/2}}
\left[2n - \frac{(\rho-1)\left[1 + \frac{\delta(\delta-1)(B+D)}{\rho(\rho-1)(A+C)}
R^{\delta-\rho}\right]}
{\left[1 + \frac{\delta(B+D)}{\rho(A+C)}R^{\delta-\rho}\right]}\right]^{-1}\, ,
\nonumber
\end{align}
and
\begin{align}
\frac{2E\sqrt{G}}{a^3} + \frac{3}{2f'(R)}\left(\frac{df'(R)}{dt}\right)^2 
= R^{\rho} \left[(A-C) \left(1 + \frac{(B-D)}{(A-C)}R^{\delta-\rho}\right) 
+ \frac{\rho(A+C)(\rho-1)^2
\left[1 + \frac{\delta(\delta-1)(B+D)}{\rho(\rho-1)(A+C)}
R^{\delta-\rho}\right]^2} {4n(4n-1)\left[1 
+ \frac{\delta(B+D)}{\rho(A+C)}R^{\delta-\rho}\right]}\right]\, .
\nonumber
\end{align}
Consequently, $z(t)$ takes the following form
\begin{align}
z(t) = \sqrt{3}a_0^n \left[12n(4n-1)\right]^n \frac{\sqrt{P(R)}}{Q(R)} 
\frac{1}{R^{n + 1/2 - \rho/2}}\, ,
\label{sp4}
\end{align}
where $P(R)$ and $Q(R)$ are defined as follows
\begin{align}
P(R) = \left[4n(4n-1)(A-C) \left(1 + \frac{(B-D)}{(A-C)}R^{\delta-\rho}\right) 
+ \frac{\rho(A+C)(\rho-1)^2
\left[1 + \frac{\delta(\delta-1)(B+D)}{\rho(\rho-1)(A+C)}R^{\delta-\rho}\right]^2}
{\left[1 + \frac{\delta(B+D)}{\rho(A+C)}R^{\delta-\rho}\right]}\right]\, ,
\label{P}
\end{align}
and
\begin{align}
Q(R) = \left[2n - \frac{(\rho-1)\left[1 + \frac{\delta(\delta-1)(B+D)}{\rho(\rho-1)(A+C)}
R^{\delta-\rho}\right]}
{\left[1 + \frac{\delta(B+D)}{\rho(A+C)}R^{\delta-\rho}\right]}\right]\, .
\label{Q}
\end{align}
Before moving further,  we check at this stage whether $Q(R)$ goes to zero or, equivalently, $z(t) \rightarrow \infty$ at some  time value. It is important 
to examine this, because as we will see, the Mukhanov-Sasaki equation (which is essential to determine the observable quantities) has a term containing 
$1/z(t)$ and, moreover, the Mukhanov variable ($v=z\Psi$) diverges at the point where $z(t)$ goes to infinity. As mentioned earlier, perturbations are
generated in the low curvature regime deeply in the contracting era and thus the above expression 
of $Q$ can be simplified, as follows
\begin{eqnarray}
 Q(R) = (2n - \rho + 1) - \frac{\delta(\delta - \rho)(B+D)}{\rho(A+C)}R^{\delta-\rho},
 \label{report1}
\end{eqnarray}
where $\rho = \frac{1}{4}\left[3 - 2n - \sqrt{1 + 4n(5+n)}\right]$, $\delta = \frac{1}{4}\left[3 - 2n + \sqrt{1 + 4n(5+n)}\right]$ and 
$A$, $B$ are two integration constants. Further, recall, 
the explicit expressions of $C$ and $D$ (see Eq.~(\ref{low curvature sol2})) are 
\begin{eqnarray}
 C = A\bigg[\frac{(6n-1)}{3n(4n-1)}\rho - \frac{(3-4n)}{3n(4n-1)}\rho(\rho-1) - \frac{2}{3n(4n-1)}\rho(\rho-1)(\rho-2)\bigg]\nonumber\\
 D = B\bigg[\frac{(6n-1)}{3n(4n-1)}\delta - \frac{(3-4n)}{3n(4n-1)}\delta(\delta-1) - \frac{2}{3n(4n-1)}\delta(\delta-1)(\delta-2)\bigg].
 \nonumber
\end{eqnarray}
Putting these expressions of $C$ and $D$ into Eq.~(\ref{report1})
\begin{eqnarray}
 Q(R) = (2n - \rho + 1) - 
 \frac{B\delta(\delta - \rho)
 \bigg(1 + \frac{(6n-1)}{3n(4n-1)}\delta - \frac{(3-4n)}{3n(4n-1)}\delta(\delta-1) - \frac{2}{3n(4n-1)}\delta(\delta-1)(\delta-2)\bigg)}
 {A\rho\bigg(1 + \frac{(6n-1)}{3n(4n-1)}\rho - \frac{(3-4n)}{3n(4n-1)}\rho(\rho-1) - \frac{2}{3n(4n-1)}\rho(\rho-1)(\rho-2)\bigg)}R^{\delta-\rho},
 \label{report2}
\end{eqnarray}
and using the forms of $\rho$ and $\delta$ (in terms of $n$), it can be checked that the above expression for $Q$ is a strictly positive definite quantity 
 for $n > \frac{1}{4}$. Moreover, we will show in the later sections that the observable quantities are compatible 
with Planck observations for the parametric regime $0.27 \lesssim n \lesssim 0.40$ (i.e for $n > 1/4$). Therefore, $Q(t)$ does not hit zero ,
or equivalently $z(t)$ does not tend to infinity, for the parametric values which are consistent with the Planck observations.

Eq.~(\ref{sp2}) clearly indicates that $\Psi(t,\vec{x})$ is not
canonically normalized and, to this end, we introduce the well-known
Mukhanov-Sasaki variable, as $v = z\Re$ ($= z\Psi$, since we are
working in the comoving gauge). The corresponding Fourier mode of
the Mukhanov-Sasaki variable satisfies
\begin{align}
\frac{d^2v_k}{d\tau^2} + \left(k^2 - \frac{1}{z(\tau)}\frac{d^2z}{d\tau^2}\right)v_k(\tau) 
= 0 \, ,
\label{sp5}
\end{align}
where $\tau = \int dt/a(t)$ is the conformal time and $v_k(\tau)$
 the Fourier transform of the variable of $v(t,\vec{x})$ for the
$k$th mode. Eq.~(\ref{sp5}) is quite hard to solve analytically, in
general, since the function $z$ depends on the background
dynamics. However, the equation can be solved analytically in the
regime $R/a_0 \ll 1$, as we now show. The conformal time ($\tau$)
is related to the cosmic time ($t$) as $\tau = \int
\frac{dt}{a(t)} = \frac{1}{a_0^n(1-2n)} t^{1-2n}$, for $n \neq
1/2$, however we will show that the observable quantities are
compatible with Planck data \cite{Akrami:2018odb} for $n < 1/2$
and thus we can safely work with the aforementioned expression of
$\tau = \tau(t)$. Using this, we can express the Ricci scalar as a
function of the conformal time
\begin{align}
R(\tau)= \frac{12n(4n-1)}{t^2} 
=\frac{12n(4n-1)}{\left[a_0^n(1-2n)\right]^{2/(1-2n)}}
\frac{1}{\tau^{2/(1-2n)}}\, .
\label{sp6}
\end{align}
Having this in mind, along with Eq.~(\ref{sp4}), we can express
$z$ in terms of $\tau$, as follows
\begin{align}
z(\tau) = \sqrt{3}a_0^n \left[12n(4n-1)\right]^n \frac{\sqrt{P(\tau)}}{Q(\tau)}
\tau^{\frac{2n+1-\rho}{1-2n}}\, .
\label{sp67}
\end{align}
The above form of $z = z(\tau)$ yields the expression of
$\frac{1}{z}\frac{d^2z}{d\tau^2}$ that is essential for the
Mukhanov equation
\begin{align}
\frac{1}{z}\frac{d^2z}{d\tau^2}=&\frac{\xi(\xi-1)}{\tau^2} 
\left[1 + \frac{2(\delta-\rho)}{(\xi-1)}R^{\delta-\rho} \right. \nonumber\\
&\left. \times \left(\frac{\delta(\rho-\delta)(B+D)}{\rho(A+C)(2n-\rho+1)} 
+ \frac{\delta(1-\rho)(1+\rho-2\delta)(B+D) + 4(B-D)n + 16(B-D)n^2}
{\rho(1-\rho)^2(A+C) + 4(A-C)n + 16(A-C)n^2}\right)\right] \, ,
\label{sp7}
\end{align}
with $\xi = \frac{(2n+1-\rho)}{(1-2n)}$. Recall $\rho =
\frac{1}{4}\left[3 - 2n - \sqrt{1 + 4n(5+n)}\right]$ and $\delta =
\frac{1}{4}\left[3 - 2n + \sqrt{1 + 4n(5+n)}\right]$, which clearly
indicate that $\delta - \rho$ is a positive quantity. Thus, the
term within parenthesis in Eq.~(\ref{sp7}) can be safely
considered to be small in the low-curvature regime $R/a_0 \ll 1$.
As a result, $\frac{1}{z}\frac{d^2z}{d\tau^2}$ becomes
proportional to $1/\tau^2$ i.e., $\frac{1}{z}\frac{d^2z}{d\tau^2} =
\sigma/\tau^2$, with
\begin{align}
\sigma = \xi(\xi-1)&\left[1 + \frac{2(\delta-\rho)}{(r-1)}R^{\delta-\rho}
\right. \nonumber\\
& \left. \times \left(\frac{\delta(\rho-\delta)(B+D)}{\rho(A+C)(2n-\rho+1)} 
+ \frac{\delta(1-\rho)(1+\rho-2\delta)(B+D) + 4(B-D)n + 16(B-D)n^2}
{\rho(1-\rho)^2(A+C) + 4(A-C)n + 16(A-C)n^2}\right)\right]\, ,
\label{spnew}
\end{align}
which is approximately a constant in this era, when the primordial
perturbation modes are generated deeply inside the Hubble radius. In
fact, and in conjunction with the fact that $c_s^2 = 1$, the
Mukhanov equation can be solved, as follows
\begin{align}
v(k,\tau) = \frac{\sqrt{\pi|\tau|}}{2} \left[c_1(k)H_{\nu}^{(1)}(k|\tau|) +
c_2(k)H_{\nu}^{(2)}(k|\tau|)\right]\, ,
\label{sp8}
\end{align}
with $\nu = \sqrt{\sigma + \frac{1}{4}}$, and $c_1$ and $c_2$ are
integration constants. Assuming the Bunch-Davies vacuum initially,
these integration constants become $c_1 = 0$ and $c_2 =1$, 
respectively. Having the solution of $v_k(\tau)$ at hand, next we
proceed to evaluate the power spectrum (defined for the
Bunch-Davies vacuum state) corresponding to the $k$-th scalar
perturbation mode, which is defined as 
\begin{align}
P_{\Psi}(k,\tau) = \frac{k^3}{2\pi^2}\left|\Psi_k(\tau)\right|^2 
= \frac{k^3}{2\pi^2}\left|\frac{v_k(\tau)}{z(\tau)}\right|^2\, .
\label{sp9}
\end{align}
In the superhorizon limit, using the mode solution in 
Eq.~(\ref{sp8}), we get
\begin{align}
P_{\Psi}(k,\tau) = \left[\frac{1}{2\pi}\frac{1}{z|\tau|}
\frac{\Gamma(\nu)}{\Gamma(3/2)}\right]^2 \left(\frac{k|\tau|}{2}\right)^{3 - 2\nu}\, .
\label{sp10}
\end{align}
By using Eq.~(\ref{sp10}), we can determine the observable
quantities, as the spectral index of the primordial curvature
perturbations and the running of the spectral index. Before proceeding
to calculate these observable quantities, we will consider first
the tensor power spectrum, which is necessary for evaluating the
tensor-to-scalar ratio.


\subsection{Tensor perturbations}

Let us now focus on the tensor perturbations, which for the FRW metric background are noted by $h_{ij}(t,\vec{x})$ and defined as 
\begin{align}
ds^2 = -dt^2 + a(t)^2\left(\delta_{ij} + h_{ij}\right)dx^idx^j\, ,
\label{tp1}
\end{align}
A tensor
perturbation is itself a gauge invariant quantity, and the tensor
perturbed action up to quadratic order is given by
\begin{align}
\delta S_{h} = \int dt d^3\vec{x} a(t) z_T(t)^2\left[\dot{h_{ij}}\dot{h^{ij}} 
 - \frac{1}{a^2}\left(\partial_kh_{ij}\right)^2\right]\, ,
\label{tp2}
\end{align}
where $z_T(t)$ is 
\begin{align}
z_T(t) = a\sqrt{f'(R)}\, ,
\label{tp3}
\end{align}
Therefore, the speed of the tensor perturbation propagation is
$c_T^2 = 1$. Similar to scalar perturbations, the Mukhanov-Sasaki variable for
tensor perturbation is defined as $(v_T)_{ij} = z_T~h_{ij}$, which,
upon performing the Fourier transformation, satisfies the equation
\begin{align}
\frac{d^2v_T(k,\tau)}{d\tau^2} 
+ \left(k^2 - \frac{1}{z_T(\tau)}\frac{d^2z_T}{d\tau^2}\right)v_T(k,\tau) 
= 0\, .
\label{tp4}
\end{align}
Using Eq.~(\ref{tp3}), along with the condition $R/a_0 \ll 1$,
we evaluate $z_T(\tau)$ and
$\frac{1}{z_T(\tau)}\frac{d^2z_T}{d\tau^2}$, and these read
\begin{align}
z_T(\tau) = a_0^n \left[12n(4n-1)\right]^n S(\tau)
\tau^{\frac{2n+1-\rho}{1-2n}}\, ,
\label{tpnew}
\end{align}
and
\begin{align}
\frac{1}{z_T}\frac{d^2z_T}{d\tau^2} = \frac{\xi(\xi-1)}{\tau^2} 
\left[1 - \frac{2\delta(\delta-\rho)(B+D)}{(r-1)\rho(A+C)}
R^{\delta-\rho}\right]\, ,
\label{tp6}
\end{align}
respectively, where $S(R(\tau)) = \sqrt{\frac{\rho(A+C)}{2}}
\left[1 + \frac{\delta(B+D)}{\rho(A+C)}R^{\delta-\rho}\right]^{1/2}$ and also we used $R=R(\tau)$ from Eq.~(\ref{sp6}). Due to the fact that
$\delta-\rho$ is positive, the variation of the term in 
parenthesis in Eq.~(\ref{tp6}) can be regarded to be small in the
low-curvature regime, and thus
$\frac{1}{z_T}\frac{d^2z_T}{d\tau^2}$ becomes proportional to
$1/\tau^2$ that is $\frac{1}{z_T}\frac{d^2z_T}{d\tau^2} =
\sigma_T/\tau^2$, with
\begin{align}
\sigma_T = \xi(\xi-1) \left[1 - \frac{2\delta(\delta-\rho)(B+D)}{(r-1)\rho(A+C)}
R^{\delta-\rho}\right]\, ,
\label{tp9}
\end{align}
and recall that $\xi = \frac{(2n+1-\rho)}{(1-2n)}$. The above
expressions yield the tensor power spectrum, defined with the initial
state being the Bunch-Davies vacuum, so we have
\begin{align}
P_{h}(k,\tau) = 2\left[\frac{1}{2\pi}\frac{1}{z|\tau|}
\frac{\Gamma(\nu_T)}{\Gamma(3/2)}\right]^2 \left(\frac{k|\tau|}{2}
\right)^{3 - 2\nu_T}\, .
\label{tp10}
\end{align}
The factor $2$ arises due to the two polarization modes of the
gravity wave, and $\nu_T = \sqrt{\sigma_T + \frac{1}{4}}$, where
$\sigma_T$ is defined in Eq.~(\ref{tp9}).

\begin{figure}[!h]
\begin{center}
 \centering
 \includegraphics[width=3.5in,height=2.0in]{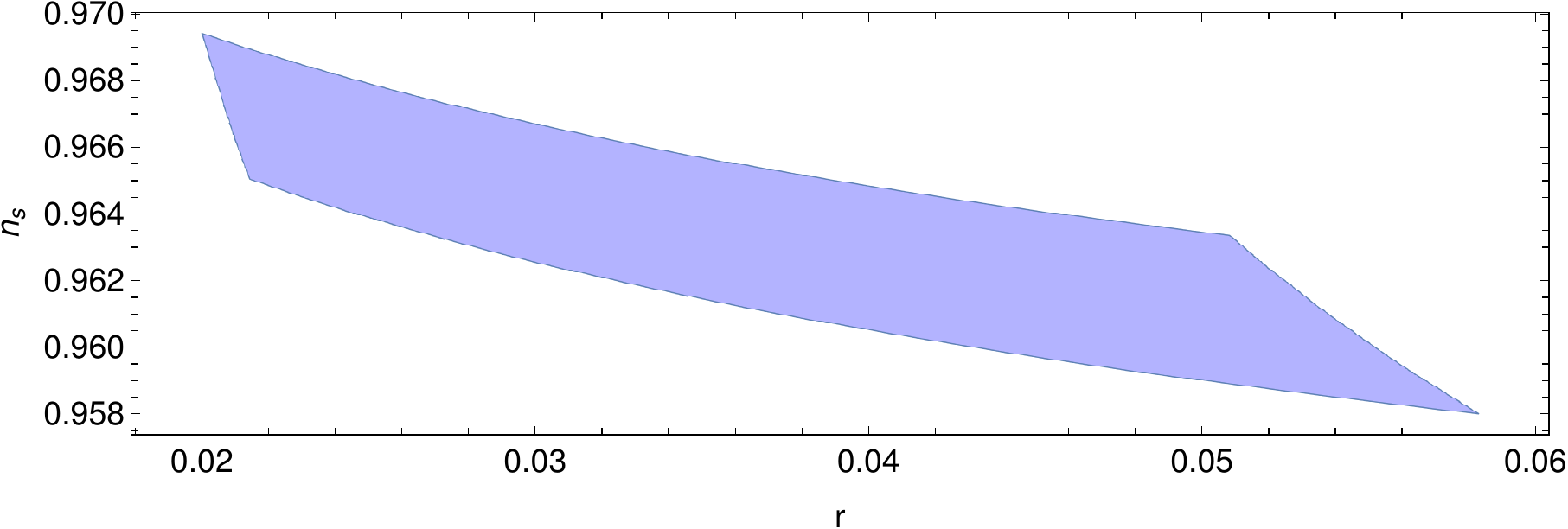}
 \caption{Parametric plot of $n_s$ vs $r$ for
 $0.01 \leq \frac{R_h}{a_0} \leq 0.07$ and $0.27 \lesssim n \lesssim 0.40$.}
 \label{plot1}
\end{center}
\end{figure}

\begin{figure}[!h]
\begin{center}
 \centering
 \includegraphics[width=3.5in,height=2.0in]{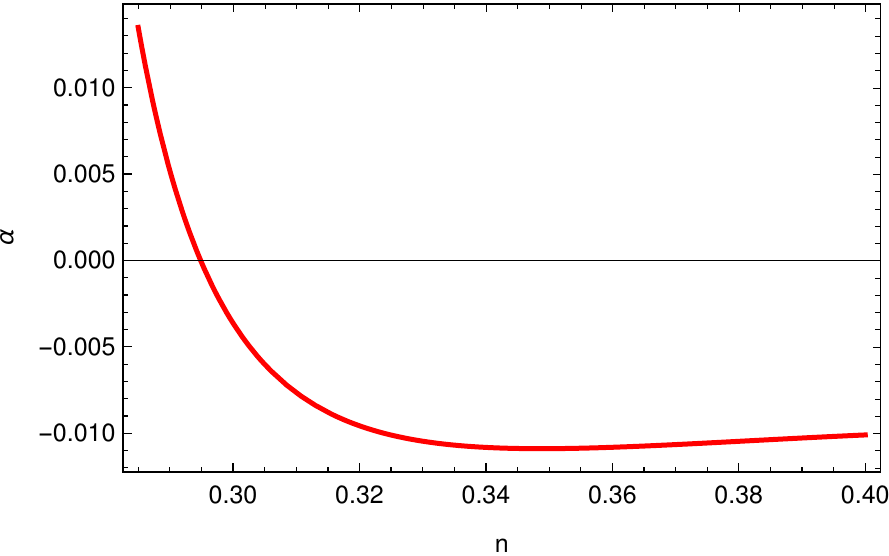}
 \caption{Parametric plot of $\alpha$ vs $n$ for
 $\frac{R_h}{a_0} = 0.05$ and $0.26 \lesssim n \lesssim 0.40$.}
 \label{plot2}
\end{center}
\end{figure}

Now, we can explicitly confront the model at hand with the latest
Planck observational data \cite{Akrami:2018odb}, so we shall
calculate the spectral index of the primordial curvature
perturbations $n_s$ and the tensor-to-scalar ratio $r$, which are
defined as follows
\begin{align}
n_s = 1 + \left. \frac{\partial \ln{P_{\Psi}}}{\partial
\ln{k}}\right|_{\tau=\tau_h} \, , \quad 
r = \left. \frac{P_{\Psi}(k,\tau)}{P_{h}(k,\tau)}\right|_{\tau=\tau_h}\, .
\label{obs1}
\end{align}
Eqs.~(\ref{sp10}) and (\ref{tp10}) immediately lead to the
explicitly form of $n_s$ and $r$, as 
\begin{align}
n_s = 4 - \sqrt{1 + 4\sigma} \, , \quad 
r = 2\left[\frac{z(\tau)}{z_T(\tau)}\right]^2_{\tau = \tau_h}\, ,
\label{obs2}
\end{align}
where $\sigma$, $z(\tau)$ and $z_T(\tau)$ are given in 
Eqs.~(\ref{spnew}), (\ref{sp6}), and (\ref{tpnew}), respectively. As it
is evident from the above equations, $n_s$ and $r$ are evaluated
at the time of  exit from the horizon, when $k=aH$ or, equivalently, at $\tau
= \tau_h$. It may be noticed that $n_s$ and $r$ depend on the
dimensionless parameters $\frac{R_h}{a_0}$ and $n$ with $R_h =
R(\tau_h)$. We can now directly confront the spectral index and
the tensor-to-scalar ratio with the Planck 2018 constraints
\cite{Akrami:2018odb}, which constrain the observational indices
as follows
\begin{equation}
\label{planckconstraints}
n_s = 0.9649 \pm 0.0042\, , \quad r < 0.064\, .
\end{equation}
For the model at hand, $n_s$ and $r$ are within the Planck
constraints for the following ranges of parameter values: $0.01
\leq \frac{R_h}{a_0} \leq 0.07$ and $0.27 \lesssim n \lesssim
0.40$. This behavior is depicted in Fig.~\ref{plot1}. The
viable range of $R_h/a_0$ is in agreement with the low-curvature
condition $R/a_0 \ll 1$ that we have considered in our
calculations. Moreover, the range of the parameter $n$ clearly
indicates that the matter bounce scenario, for which $n=1/3$, is
well described by the Lagrange multiplier $F(R)$ gravity model with the holonomy corrections (though 
the holonomy corrections may be disregarded in calculating the observable quantities of the low curvature regime, as mentioned earlier). 
At this stage it is worth mentioning 
that in scalar-tensor theory (with an exponential scalar
potential), the matter bounce scenario is not consistent with
the Planck observations. Moreover, the matter bounce scenario also does
not fit well even in the standard $F(R)$ gravity, as we confirmed in an earlier paper \cite{tp_bouncing1}. 
However, here we show that for the 
Lagrange multiplier generalized $F(R)$ gravity, the matter bounce
may indeed be considered as a good bouncing model, which allows the
simultaneous compatibility of $n_s$ and $r$ with observations. Here it may be mentioned that an analogue study of matter bounce cosmology 
can be found in \cite{Cai:2011tc} within another type of modified gravity, in particular $F(T)$ gravity, where the power spectrum becomes 
nearly scale invariant but suffers from an over large tensor-to-scalar ratio. However, in the present paper, we showed that within holonomy corrected 
Lagrange multiplier $F(R)$ gravity, the matter bounce scenario gives rise to a nearly scale invariant power spectrum and also the tensor-to-scalar ratio 
lies within observational bound.\\
The running of the spectral index is defined as
\begin{align}
 \alpha = \left. \frac{dn_s}{d\ln{k}} \right|_{\tau=\tau_h}\, ,
 \label{obs3}
\end{align}
and this is constrained by the Planck 2018 results, as $\alpha =
-0.0085 \pm 0.0073$. Thus, it is also important to calculate the
running of the spectral index before concluding the viability of a
model. By using the expression of $\sigma$ (see Eq.~(\ref{spnew}))
and $R = R(\tau)$ (see Eq.~(\ref{sp6})), we get
\begin{align}
\alpha=&\frac{4\xi(\delta-\rho)^2}{\sqrt{1 + 4\xi(\xi-1)}}R_h^{\delta-\rho} 
\nonumber\\
& \times \left(\frac{\delta(\rho-\delta)(B+D)}{\rho(A+C)(2n-\rho+1)} 
+ \frac{\delta(1-\rho)(1+\rho-2\delta)(B+D) + 4(B-D)n + 16(B-D)n^2}
{\rho(1-\rho)^2(A+C) + 4(A-C)n + 16(A-C)n^2}\right)\, .
\label{obs4}
\end{align}
To arrive at the above result, we use the horizon crossing
relation of the $k$-th mode $k = aH$ to determine
$\frac{d|\tau|}{d\ln{k}} = -|\tau|$ i.e., the horizon exit time
$|\tau|$ increases as the momentum of the perturbation mode
decreases, as expected. Eq.~(\ref{obs4}) indicates that, similarly
to $n_s$ and $r$, the running index ($\alpha$) also depends on the
parameters $R_h/a_0$ and $n$. Taking $R_h/a_0 = 0.05$, we have produced a
plot of $\alpha$ with respect to $n$, in Fig.~\ref{plot2}.
As can be seen in Fig.~\ref{plot2}, the parameter $\alpha$
takes negative values, crossing  zero about $n \simeq 0.30$.
Thus $\alpha$ lies within the Planck constraint for $0.30 \lesssim
n \lesssim 0.40$, which includes the matter bounce scenario. For
the Lagrange multiplier generalized $F(R)$ gravity model, we
showed that the pure matter bounce scenario, as well as the quasi
matter bounce scenario, are both consistent with the Planck
observations. Therefore, generalized $F(R)$ gravity in terms of the 
Lagrange multiplier has a richer phenomenology, in comparison with
scalar-tensor or standard $F(R)$ gravity models, which fail to
describe in a viable way these two bouncing cosmology scenarios.



\section{Energy conditions}\label{sec_energy}

\begin{figure}[!h]
\begin{center}
 \centering
 \includegraphics[width=3.0in,height=2.0in]{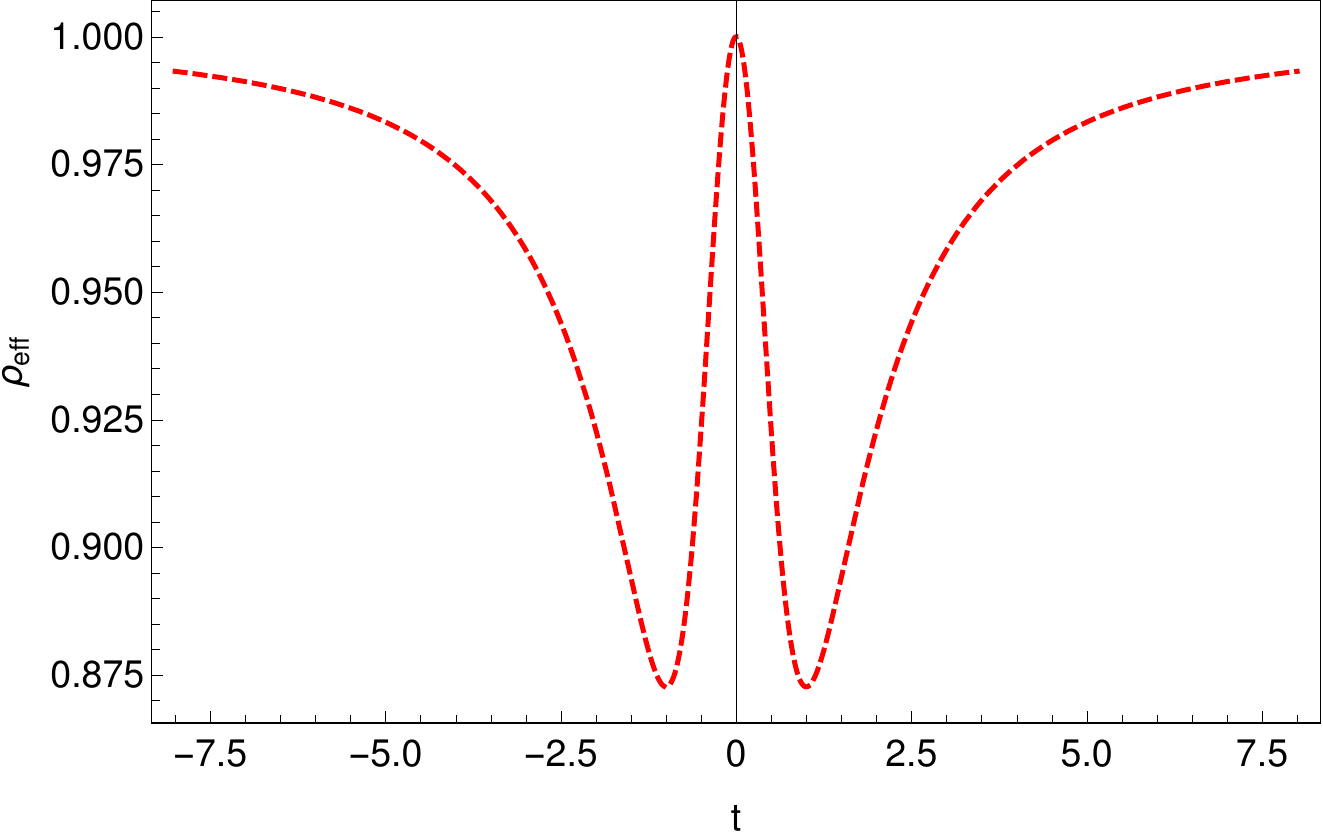}
 \includegraphics[width=3.0in,height=2.0in]{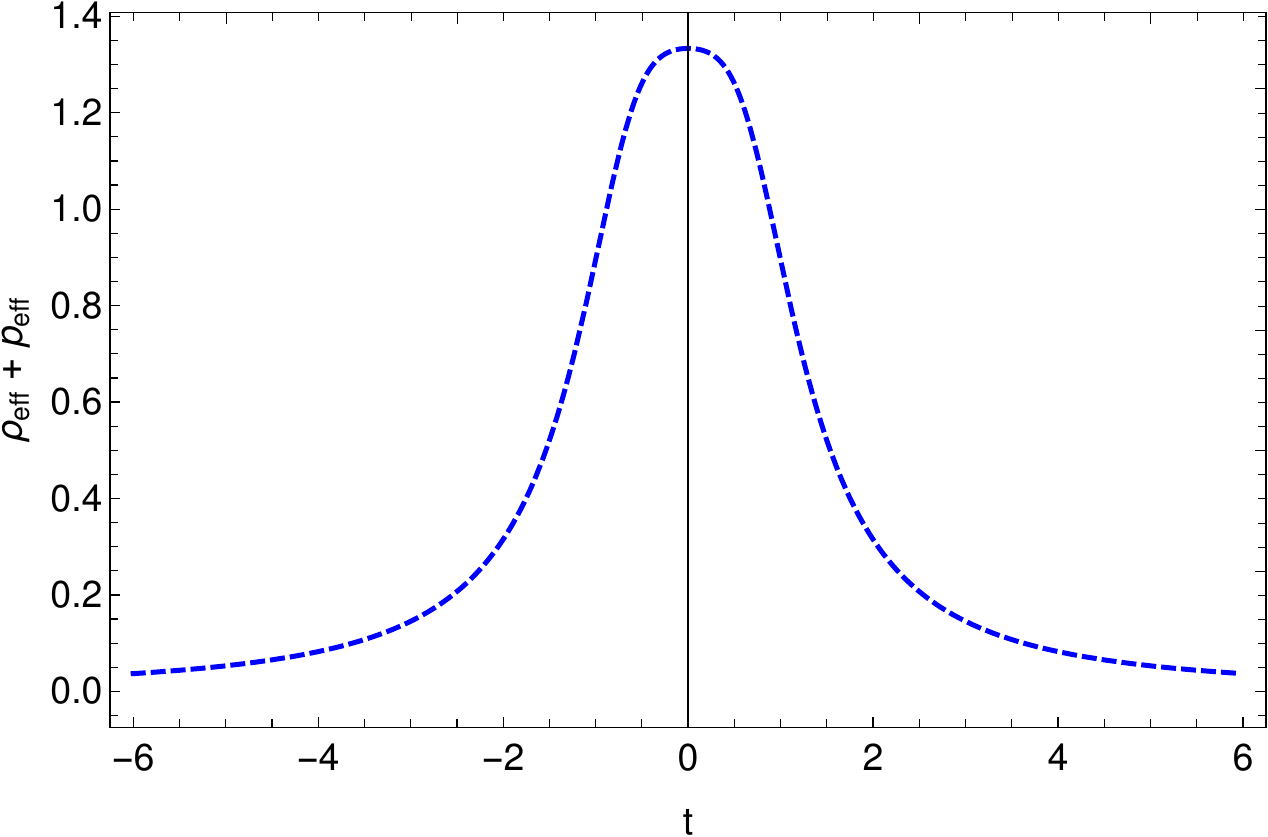}
 \caption{$Left~part$: $\rho_\mathrm{eff}/\rho_c$ vs. $t$ for the weak energy condition. $Right~part$ : $\rho_\mathrm{eff} + p_\mathrm{eff}$ vs.$t$ for the
  null energy condition. In both cases, we take $k\gamma = 1$, $a_0 = 1$ (in reduced Planck
units) and $n = 1/3$.}
 \label{plot_energy}
\end{center}
\end{figure}

A crucial drawback in most  bouncing models is the violation
of the null energy condition near the bounce at which the bouncing phenomena is realized. However in the present context, we deal 
with the matter or quasi-matter bounce in the backdrop of the 
Lagrange multiplier $F(R)$ gravity model in presence of holonomy corrections. And it does happen that such holonomy corrections may become significant 
about the bouncing point (where the spacetime curvature is large compared to the one at present) and thus may play an important 
role in restoring the energy conditions. Keeping this in mind, here we check the energy conditions in Lagrange multiplier F(R) gravity 
 with holonomy corrections, where the effective energy density $\rho_\mathrm{eff}$ and pressure
$p_\mathrm{eff}$ can be determined from 
\begin{eqnarray}
 H^2&=&\bigg(\frac{1-3A/4\rho_c}{3}\bigg)\bigg[\frac{R}{2} - \frac{1}{F_R + \frac{EG_R}{a^3\sqrt{G}}}\bigg(\frac{F}{2} + \frac{E\sqrt{G}}{a^3} 
 + \frac{3H}{\sqrt{1-3A/4\rho_c}}\frac{d}{dt}\big(F_R + \frac{EG_R}{a^3\sqrt{G}}\big)\bigg)\bigg]\nonumber\\
 &=&\frac{\rho_\mathrm{eff}}{3} \bigg[1 - \frac{\rho_\mathrm{eff}}{\rho_c}\bigg]
 \label{ec1}
\end{eqnarray}
and
\begin{eqnarray}
 \dot{H}&=&\frac{1}{2f_R}\bigg(\bigg[\big(1-3A/4\rho_c\big) - 2Hk\gamma f_R\frac{\sqrt{3A/4\rho_c}}{\sqrt{1-3A/4\rho_c}}\bigg]
 \bigg[-\frac{F}{2} - f_R\big[-\frac{R}{2} + \frac{3H^2}{1-3A/4\rho_c}\big]\bigg] - 2H\frac{df_R}{dt} - \sqrt{1-3A/4\rho_c}\frac{d^2f_R}{dt^2}\bigg)\nonumber\\
 &=&-\frac{1}{2}\big(\rho_\mathrm{eff} + p_\mathrm{eff}\big) \bigg[1 - \frac{2\rho_\mathrm{eff}}{\rho_c}\bigg],
 \label{ec2}
\end{eqnarray}
respectively, with $f_R = F_R + \frac{EG_R}{a^3\sqrt{G}}$. 
Eq.~(\ref{ec1}) can be simplified to $\rho_\mathrm{eff}^2 - \rho_\mathrm{eff} \rho_c + 3H^2\rho_c = 0$, which can be solved 
as $\rho_\mathrm{eff} = \frac{\rho_c}{2}\bigg[1 \pm \sqrt{1-\frac{12H^2}{\rho_c}}\bigg]$. At the bouncing point, the Hubble parameter becomes zero, 
 $H(t=0) = 0$, which immediately leads to $\rho_\mathrm{eff}^{(+)}(t=0) = \rho_c$ and $\rho_\mathrm{eff}^{(-)}(t=0) = 0$ (we call  $'+'$ and $'-'$ the
branch solutions of $\rho_\mathrm{eff}$). Recall that, without the holonomy improvement, the effective energy density 
goes to zero at the bouncing point, whereby the $'+'$ branch solution comes just due to the presence of holonomy corrections. Considering that 
such holonomy corrections have an effect on the evolution of $\rho_\mathrm{eff}$ near the bounce, we take the $'+'$ branch solution; otherwise 
the holonomy corrections would have no effect on the effective energy density even at the bouncing point. Thus the evolution of $\rho_\mathrm{eff}$ is given by 
\begin{eqnarray}
 \rho_\mathrm{eff} = \frac{\rho_c}{2}\bigg[1 + \sqrt{1-\frac{12H^2}{\rho_c}}\bigg]
 \label{ec3}
\end{eqnarray}
Putting this solution into Eq.~(\ref{ec2}), we get
\begin{eqnarray}
p_\mathrm{eff} = \frac{2\dot{H}}{\sqrt{1-\frac{12H^2}{\rho_c}}} - \frac{\rho_c}{2}\big[1 + \sqrt{1-\frac{12H^2}{\rho_c}}\big],
\label{ec4}
\end{eqnarray}
and using the expressions of the Hubble parameter in the present context ($H = \frac{dot{a}}{a}$ with $a(t) = \big(a_0t^2 + 1\big)^n$), we give the plots of 
$\rho_\mathrm{eff}/\rho_c$ and $\rho_\mathrm{eff} + p_\mathrm{eff}$ (with respect to the cosmic time) as the left and right plots of Fig.~\ref{plot_energy}, 
respectively, for $n = 1/3$, $k\gamma = 1$, $a_0 = 1$ (in reduced Planck
units).

As it can be observed from Fig.~\ref{plot_energy} that $\rho_\mathrm{eff} + p_\mathrm{eff}$ asymptotically goes to zero, as expected since $\dot{H}$ vanishes for 
$t \rightarrow \pm \infty$. Moreover Fig.~\ref{plot_energy} clearly demonstrates that both the weak and the null energy 
conditions are satisfied in the holonomy corrected version of Lagrange multiplier $F(R)$ gravity, unlike in the usual Friedmann case where 
the energy conditions are generally violated near the bouncing point. It is worth mentioning that in the usual Friedmann cosmology, 
the gravitational equations 
come as $H^2 = \frac{\rho_\mathrm{eff}}{3}$ and $\dot{H} = -\frac{1}{2}\big[\rho_\mathrm{eff} + p_\mathrm{eff}\big]$ respectively and thus 
at the bouncing point (where $H = 0$ and $\dot{H} > 0$), $\rho_\mathrm{eff} + p_\mathrm{eff}$ becomes less than zero, which implies the violation 
of the energy conditions. However in presence of the holonomy improvements, there are extra terms 
in the equation of motion, as for example $\dot{H} = -\frac{1}{2}\big[\rho_\mathrm{eff} + p_\mathrm{eff}\big]\big[1 - 2\rho_\mathrm{eff}/\rho_c\big]$, 
and thus, at the bouncing point (where $\rho_\mathrm{eff} = \rho_c$ and $\dot{H} > 0$), $\rho_\mathrm{eff} + p_\mathrm{eff}$ becomes positive, 
so that the energy conditions are restored. Therefore, it is clear that the holonomy improvement plays a significant 
role to rescue the energy conditions for the matter (or quasi-matter) bounce scenario.

\begin{figure}[!h]
\begin{center}
 \centering
 \includegraphics[width=3.5in,height=2.0in]{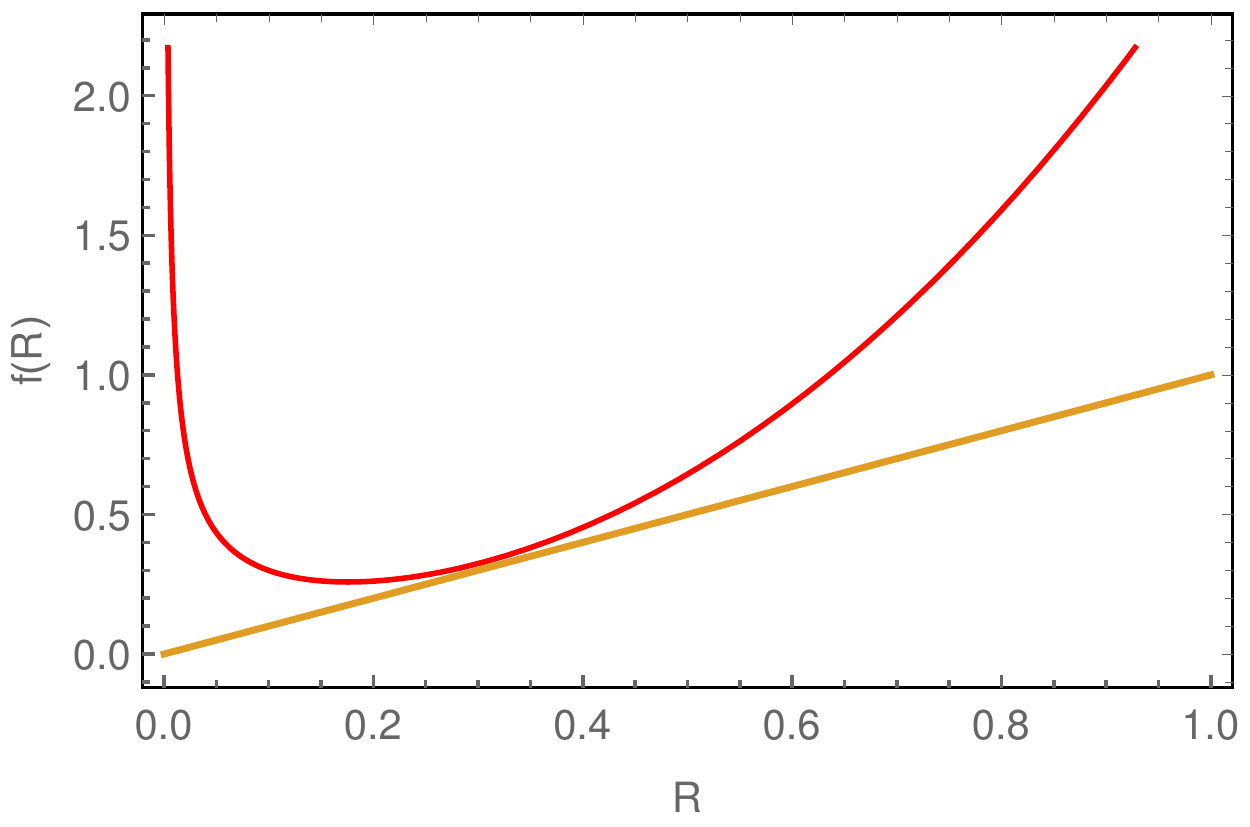}
 \caption{Numerical solution of $f = f(R)$ with $R$ being the independent variable. We take $a_0 = 1$ (in reduced Planck unit), $n = 1/3$ and 
 $k\gamma = 1$. The straight line is a plot of the Ricci scalar.}
 \label{plot_numerical}
\end{center}
\end{figure}

Before concluding, we determine the form of the effective f(R) in the whole curvature regime. This has to be done numerically, owing to the complicated equations 
of motion. In a previous Section [], we noticed that in the low-curvature regime, $f(R)$ goes as $f(R) \propto R^{\rho}$. Recall, 
$\rho = \frac{1}{4}\left[3-2n-\sqrt{1+4n(5+n)}\right]$ which is negative 
for $n > 0.25$, and as shown earlier, the present model is 
consistent with the Planck results for $0.27 \lesssim n \lesssim 0.4$, hence $\rho$ is negative, in order to ensure the viability of the
model. Therefore, it is clear that in the low-curvature regime $f(R)$ is 
proportional to an inverse power of Ricci scalar $\propto R^{-|\rho|}$, using such form of $f(R)$ as 
boundary condition along with the expression 
$R(t) = 12n \left[\frac{(4n-1)t^2 + 1/a_0}{\left(t^2 + 1/a_0\right)^2}\right]$, we solve Eqs.~(\ref{Mh_final1}) and (\ref{Mh_final2}) numerically, 
with the cosmic time $t$ being the independent variable. Moreover, $a_0$, $n$ and $k\gamma$ are taken as $a_0 = 1$ (in reduced Planck units), $n = 1/3$ 
and $k\gamma = 1$, respectively; so in effect we consider the matter bounce 
scenario. However, it may be mentioned that the $n = 1/3$ case 
makes the model consistent with the Planck 2018 constraints, as 
confirmed in the previous section. The numerical solution of $f(R)$ in terms of $R$ is obtained by using the expression 
$R(t) = 12n \left[\frac{(4n-1)t^2 + 1/a_0}{\left(t^2 + 1/a_0\right)^2}\right]$ and is presented in Fig.~\ref{plot_numerical}. We also give a plot of the
Ricci scalar in the same diagram, in order to make a comparison of the effective f(R) in the present context with the one corresponding to Einstein's gravity.

It is clear from Fig.~\ref{plot_numerical} that $f(R)$ decreases
in the regime $R/a_0 \ll 1$ (i.e $f(R) \sim R^{\rho}$), while in $R/a_0 \gtrsim 1$, $f(R)$
increases as a function of the Ricci scalar (i.e $f(R) \sim R^{\delta}$), and in the intermediate scale, 
$f(R)$ matches Einstein's gravity. Having such form of f(R), it is important to 
explore whether this form of $f(R)$ passes the astrophysical tests in the low curvature regime and, on the other hand, 
gives rise to an accelerating universe at late time.\\
An example of the tests of infrared instability is matter instabilities, which are related to the fact that the
spherical body solution in general relativity may not be a
solution in the modified gravity theory considered. Matter instabilities may
appear when the energy density or the curvature is large compared
with the average density or curvature in the universe, as is the
case inside of a planet. Following \cite{Nojiri:2010wj}, we
immediately write the potential ($U(R_b)$, with $R_b$ the
perturbed Ricci scalar) for the perturbed Ricci curvature over
Einstein gravity, as
\begin{eqnarray}
 U(R_b) = \frac{R_b}{3} - \frac{f^{(1)}(R_b)f^{(3)}(R_b)R_b}{3f^{(2)}(R_b)^2} - \frac{f^{(1)}(R_b)}{3f^{(2)}(R_b)} +
 \frac{2f(R_b)f^{(3)}(R_b)}{3f^{(2)}(R_b)^2} - \frac{f^{(3)}(R_b)R_b}{3f^{(2)}(R_b)^2},
 \label{astro_test1}
\end{eqnarray}
where we consider $f(R) \sim R^{\rho}$ in low curvature regime and we denote $d^kf(R)/dR^k = f^{(k)}(R)$. If $U(R_b) > 0$, the
perturbation grows with time and the system becomes unstable.
Recall $\rho < 0$ and $\delta > 0$ and thus the term $R^{\rho}$
dominates over $R^{\delta}$ in the low curvature regime. Thus we
can approximate $F(R) \sim R^{\rho}$ in the low curvature regime
which immediately leads to the potential, as
\begin{eqnarray}
 U(R_b) = -\frac{2(|\rho| + 2)}{9|\rho|(|\rho| + 1)}R_b + \frac{(|\rho| + 2)}{3|\rho|(|\rho| + 1)}R_b^{2-|\rho|}.
 \label{astro_test2}
\end{eqnarray}
In the low curvature regime, the first term dominates in the above expression of $U(R_b)$, and thus $U(R_b)$ becomes negative. This indicates that
the model considered here passes the matter instability test. Moreover, here it may be mentioned that for $|\rho| = 1$, the above potential 
$U(R_b)$ becomes positive, which indicates that $f(R) \sim 1/R$ (in low curvature regime) leads to an infrared instability, as also confirmed 
in \cite{Nojiri:2010wj}. However for the purpose of investigating the late time acceleration in our present model, one may check that 
$f(R) \sim R^{\rho}$ leads to a power solution of the scale factor as $a(t) = a_0 t^{-\frac{(\rho - 1)(2\rho - 1)}{(\rho - 2)}}$ 
(see \cite{Nojiri:2010wj}), recall $\rho = \frac{1}{4}\big[3-4n-\sqrt{1+4n(5+n)}\big]$. We showed in the earlier section that the viable range 
of $n$ for which the observable parameters ($n_s$, $r$ and $\alpha$) are simultaneously compatible with Planck observations is given by 
$0.30 \lesssim n \lesssim 0.40$. However this range of $n$ leads to the exponent in the solution of $a(t)$ as 
$0 < -\frac{(\rho - 1)(2\rho - 1)}{(\rho - 2)} < 1$, which in turn 
indicates a decelerating expanding universe at late time. The deceleration of the late time scale factor is also confirmed 
from the evolution of Hubble radius (the Hubble radius is increasing at the late time of the expanding phase), as explained in 
Section[\ref{sec_reconstruction}]. Thereby as per the evidences from supernova data \cite{Perlmutter:1998np,Riess:1998cb}, 
which points out that the present universe undergoes 
through an accelerating phase, our present model is unable to describe the behaviour of our universe at late time. However our results indicate 
that the present model must be combined with another cosmological scenario in the low curvature regime of the expanding phase, 
or should be modified appropriately, in order that it leads to a viable cosmology at late time. Such unification of bouncing 
with late time acceleration is an interesting issue and is expected to study in a near future work.

\section{Conclusions}
We have discussed a bouncing scenario which incorporates holonomy corrections in an F(R) gravity model with a Lagrange multiplier and where 
the Hubble parameter squared is proportional (aside from some coefficients) 
to a linear as well as to a quadratic power of the effective energy density; this differs from  the usual Friedmann case, where $H^2$ is proportional to the linear power of energy density,  only. We have specified our study to  matter or quasi-matter bounce scenarios, with the scale factor being
$a(t) = \left(a_0t^2 + 1 \right)^n$, where $n$ is a parameter of the model  and $t$  the cosmic time. For such bouncing scale factor, it was shown that, 
for $n < 1/2$, the primordial curvature perturbation modes exit the Hubble horizon (defined by $r_h = 1/aH$) at a large negative timevalue deeply in the contracting 
era (where the spacetime curvature is low as compared to the bouncing point), which in turn makes the ``low curvature approximation'' a viable one to calculate 
the power spectra of the scalar and tensor perturbations. We have determined the form of the effective f(R) theory in the low curvature regime, by using a reconstruction technique 
for the present model, which realizes the bouncing with the aforementioned scale factor. The form of f(R) leads to the explicit 
expression of the well known Mukhanov-Sasaki equation, by solving which we have determined the power spectrums of primordial 
perturbations and, correspondingly, have calculated the fundamental cosmological parameters, such as the spectral index of scalar perturbation, the tensor to scalar ratio, and 
the running spectral index. Such observational indexes are found to depend on the dimensionless  parameters of the model, as $R_h/a_0$ (with $R_h$ the Ricci 
curvature at the time of horizon exit) and $n$. It turned out that the observable quantities are 
simultaneously compatible with the Planck2018 constraints for the parameteric range $0.01 \lesssim \frac{R_h}{a_0} \lesssim 0.07$ 
and $0.27 \lesssim n \lesssim 0.40$. It should be noticed that this range of $n$ is also supported by the range $0 < n < \frac{1}{2}$, which makes 
the low-curvature approximation a perfectly reliable one for calculating the power spectra.

 Further, we have determined the explicit expressions for the effective energy density ($\rho_\mathrm{eff}$) and effective pressure ($p_\mathrm{eff}$) 
in the present holonomy improved scenario, from which we have investigated the problem of the energy conditions. Concerning this issue,
we have obtained a quite remarkable result, namely that both the weak and the null energy conditions are fulfilled, owing precisely to the presence of the holonomy modifications in our Lagrange multiplier 
F(R) gravity model. This is a significant difference with respect to the usual Friedmann case, where the energy conditions are generically violated near the bouncing point. Actually in usual Friedmann cosmology, 
the gravitational equations have the form $H^2 = \frac{\rho_\mathrm{eff}}{3}$ and $\dot{H} = -\frac{1}{2}\big[\rho_\mathrm{eff} + p_\mathrm{eff}\big]$, 
respectively, and thus 
at the bouncing point (where $H = 0$ and $\dot{H} > 0$), $\rho_\mathrm{eff} + p_\mathrm{eff}$ becomes negative, what leads to the violation 
of the energy conditions. However, because of the holonomy corrections, extra terms appear in the equation of motion, as for instance $\dot{H} = -\frac{1}{2}\big[\rho_\mathrm{eff} + p_\mathrm{eff}\big]\big[1 - 2\rho_\mathrm{eff}/\rho_c\big]$, 
and thus at the bouncing point (where $\rho_\mathrm{eff} = \rho_c$ and $\dot{H} > 0$), $\rho_\mathrm{eff} + p_\mathrm{eff}$ becomes positive, 
what shows that the energy conditions are restored. Thus, the holonomy corrected Lagrange multiplier F(R) gravity model 
discussed here is fully vindicated as a viable description for the bouncing scenario with the aforementioned scale factor.

\section*{Appendix-1}
Here we give the details of the calculation of Eq.~(\ref{appendix}). The demonstration goes as follows.

Introduce $\tilde{p}$ as $\frac{1}{2}\bigg(\frac{p_R}{f_{RR}} + \frac{p_{\lambda}}{G_R}\bigg) = \frac{V}{f_{RR}}\tilde{p}$. With this new 
defined quantity, the first and second expressions of eqn.(\ref{Mh1}) takes the following form 
\begin{eqnarray}
 \sin^2{b} = 1 - \frac{9H^2f_{RR}^2}{\tilde{p}^2}
 \label{appendix1}
\end{eqnarray}
and
\begin{eqnarray}
 2f_{RR}\dot{R} = \frac{2}{k\gamma}\sin{b} + \frac{2f_R}{3f_{RR}}\tilde{p},
 \label{appendix2}
\end{eqnarray}
respectively. On the other hand, Eq.(\ref{Mh2}) can be simplified as
\begin{eqnarray}
 \frac{1}{2}\big[Rf_R - f\big] - \frac{1}{2V}\big[-\frac{2}{k\gamma}\sin{b}\big]\bigg(\frac{p_R}{f_{RR}} + \frac{p_{\lambda}}{G_R}\bigg) 
 + \frac{f_R}{12V^2}\bigg(\frac{p_R}{f_{RR}} + \frac{p_{\lambda}}{G_R}\bigg)^2 - \frac{1}{2}\lambda \dot{\Phi}^2 = 0\nonumber\\
 \Rightarrow \frac{1}{2}\big[Rf_R - f - \lambda\dot{\Phi}^2\big] 
 - \frac{1}{2V}\big[-\frac{2}{k\gamma}\sin{b}\big] \frac{\tilde{p}}{f_{RR}} 
 + \frac{f_R}{3f_{RR}^2}\tilde{p}^2 = 0\nonumber\\
 \Rightarrow \frac{1}{2}\big[Rf_R - f - \lambda\dot{\Phi}^2\big] 
 + \frac{\tilde{p}}{f_{RR}} \bigg[2f_{RR}\dot{R} - \frac{f_R}{3f_{RR}}\tilde{p}\bigg] = 0.
 \nonumber
\end{eqnarray}
The above expression is a quadratic equation on $\tilde{p}$, which can be solved as follows
\begin{eqnarray}
 \tilde{p} = \frac{2\dot{R} \pm \sqrt{4\dot{R}^2 + \frac{2f_R}{3f_{RR}^2}\big(Rf_R - f - \lambda\dot{\Phi}^2\big)}}{\frac{2f_R}{3f_{RR}^2}}
 \label{appendix3}
\end{eqnarray}
Eq.~(\ref{appendix3}) along with Eq.~(\ref{appendix2}) lead to the awaited expression of $\sin^2{b}$ 
\begin{eqnarray}
 \sin^2{b} = \frac{3A}{4\rho_c},
 \label{appendix4}
\end{eqnarray}
with $A = 4\big(\dot{R}f_{RR}\big)^2 + \frac{2}{3}f_R\big(Rf_R - f - \lambda\dot{\Phi}^2\big)$ and $\rho_c = \frac{3}{k^2\gamma^2}$.

\section*{Acknowledgments}

EE and SDO acknowledge the support of MINECO (Spain), project FIS2016-76363-P, and  AGAUR (Catalonia, Spain) project 2017 SGR 247.
TP acknowledges the hospitality by ICE-CSIC/IEEC (Barcelona, Spain), where a part of this work was done during his visit.


\begin{thebibliography}{99}
 
\bibitem{Brandenberger:2012zb}
  R.~H.~Brandenberger,
  arXiv:1206.4196 [astro-ph.CO].



\bibitem{Brandenberger:2016vhg}
  R.~Brandenberger and P.~Peter,
  arXiv:1603.05834 [hep-th].


\bibitem{Battefeld:2014uga}
 D.~Battefeld and P.~Peter,
 Phys.\ Rept.\ {\bf 571} (2015) 1
 [arXiv:1406.2790 [astro-ph.CO]].


\bibitem{Novello:2008ra}
 M.~Novello and S.~E.~P.~Bergliaffa,
 Phys.\ Rept.\ {\bf 463} (2008) 127
 [arXiv:0802.1634 [astro-ph]].


\bibitem{Cai:2014bea}
  Y.~F.~Cai,
  Sci.\ China Phys.\ Mech.\ Astron.\  {\bf 57} (2014) 1414
  doi:10.1007/s11433-014-5512-3
  [arXiv:1405.1369 [hep-th]].



\bibitem{deHaro:2015wda}
 J.~de Haro and Y.~F.~Cai,
 Gen.\ Rel.\ Grav.\ {\bf 47} (2015) no.8, 95
 [arXiv:1502.03230 [gr-qc]].





\bibitem{Lehners:2011kr}
 J.~L.~Lehners,
 Class.\ Quant.\ Grav.\ {\bf 28} (2011) 204004
 [arXiv:1106.0172 [hep-th]].



\bibitem{Lehners:2008vx}
 J.~L.~Lehners,
 Phys.\ Rept.\ {\bf 465} (2008) 223
 [arXiv:0806.1245 [astro-ph]].

\bibitem{Cheung:2016wik}
 Y.~K.~E.~Cheung, C.~Li and J.~D.~Vergados,
 arXiv:1611.04027 [astro-ph.CO].


\bibitem{Cai:2016hea}
  Y.~F.~Cai, A.~Marciano, D.~G.~Wang and E.~Wilson-Ewing,
  Universe {\bf 3} (2016) no.1,  1
  doi:10.3390/universe3010001
  [arXiv:1610.00938 [astro-ph.CO]].





\bibitem{Cattoen:2005dx}
 C.~Cattoen and M.~Visser,
 Class.\ Quant.\ Grav.\ {\bf 22} (2005) 4913
 [gr-qc/0508045].



\bibitem{Li:2014era}
 C.~Li, R.~H.~Brandenberger and Y.~K.~E.~Cheung,
 Phys.\ Rev.\ D {\bf 90} (2014) no.12, 123535
 [arXiv:1403.5625 [gr-qc]].




\bibitem{Brizuela:2009nk}
 D.~Brizuela, G.~A.~D.~Mena Marugan and T.~Pawlowski,
 Class.\ Quant.\ Grav.\ {\bf 27} (2010) 052001
 [arXiv:0902.0697 [gr-qc]].



\bibitem{Cai:2013kja}
 Y.~F.~Cai, E.~McDonough, F.~Duplessis and R.~H.~Brandenberger,
 JCAP {\bf 1310} (2013) 024
 [arXiv:1305.5259 [hep-th]].


\bibitem{Quintin:2014oea}
 J.~Quintin, Y.~F.~Cai and R.~H.~Brandenberger,
 Phys.\ Rev.\ D {\bf 90} (2014) no.6, 063507
 [arXiv:1406.6049 [gr-qc]].


\bibitem{Cai:2013vm}
 Y.~F.~Cai, R.~Brandenberger and P.~Peter,
 Class.\ Quant.\ Grav.\ {\bf 30} (2013) 075019
 [arXiv:1301.4703 [gr-qc]].



\bibitem{Poplawski:2011jz}
 N.~J.~Poplawski,
 Phys.\ Rev.\ D {\bf 85} (2012) 107502
 [arXiv:1111.4595 [gr-qc]].




\bibitem{Koehn:2015vvy}
 M.~Koehn, J.~L.~Lehners and B.~Ovrut,
 Phys.\ Rev.\ D {\bf 93} (2016) no.10, 103501
 [arXiv:1512.03807 [hep-th]].



\bibitem{Odintsov:2015zza}
 S.~D.~Odintsov and V.~K.~Oikonomou,
 Phys.\ Rev.\ D {\bf 92} (2015) no.2, 024016
 [arXiv:1504.06866 [gr-qc]].


\bibitem{Nojiri:2016ygo}
 S.~Nojiri, S.~D.~Odintsov and V.~K.~Oikonomou,
 Phys.\ Rev.\ D {\bf 93} (2016) no.8, 084050
 [arXiv:1601.04112 [gr-qc]].



\bibitem{Oikonomou:2015qha}
 V.~K.~Oikonomou,
 Phys.\ Rev.\ D {\bf 92} (2015) no.12, 124027
 [arXiv:1509.05827 [gr-qc]].




\bibitem{Odintsov:2015ynk}
 S.~D.~Odintsov and V.~K.~Oikonomou,
 arXiv:1512.04787 [gr-qc].


\bibitem{Koehn:2013upa}
 M.~Koehn, J.~L.~Lehners and B.~A.~Ovrut,
 Phys.\ Rev.\ D {\bf 90} (2014) no.2, 025005
 [arXiv:1310.7577 [hep-th]].



\bibitem{Battarra:2014kga}
 L.~Battarra and J.~L.~Lehners,
 JCAP {\bf 1412} (2014) no.12, 023
 [arXiv:1407.4814 [hep-th]].


\bibitem{Martin:2001ue}
 J.~Martin, P.~Peter, N.~Pinto Neto and D.~J.~Schwarz,
 Phys.\ Rev.\ D {\bf 65} (2002) 123513
 [hep-th/0112128].


\bibitem{Khoury:2001wf}
 J.~Khoury, B.~A.~Ovrut, P.~J.~Steinhardt and N.~Turok,
 Phys.\ Rev.\ D {\bf 64} (2001) 123522
 [hep-th/0103239].


\bibitem{Buchbinder:2007ad}
 E.~I.~Buchbinder, J.~Khoury and B.~A.~Ovrut,
 Phys.\ Rev.\ D {\bf 76} (2007) 123503
 [hep-th/0702154].




\bibitem{Brown:2004cs}
 M.~G.~Brown, K.~Freese and W.~H.~Kinney,
 JCAP {\bf 0803} (2008) 002
 [astro-ph/0405353].



\bibitem{Hackworth:2004xb}
 J.~C.~Hackworth and E.~J.~Weinberg,
 Phys.\ Rev.\ D {\bf 71} (2005) 044014
 [hep-th/0410142].


\bibitem{Nojiri:2006ww}
 S.~Nojiri and S.~D.~Odintsov,
 Phys.\ Lett.\ B {\bf 637} (2006) 139
 [hep-th/0603062].


\bibitem{Johnson:2011aa}
 M.~C.~Johnson and J.~L.~Lehners,
 Phys.\ Rev.\ D {\bf 85} (2012) 103509
 [arXiv:1112.3360 [hep-th]].



\bibitem{Peter:2002cn}
 P.~Peter and N.~Pinto-Neto,
 Phys.\ Rev.\ D {\bf 66} (2002) 063509
 [hep-th/0203013].


\bibitem{Gasperini:2003pb}
 M.~Gasperini, M.~Giovannini and G.~Veneziano,
 Phys.\ Lett.\ B {\bf 569} (2003) 113
 [hep-th/0306113].



\bibitem{Creminelli:2004jg}
 P.~Creminelli, A.~Nicolis and M.~Zaldarriaga,
 Phys.\ Rev.\ D {\bf 71} (2005) 063505
 [hep-th/0411270].


\bibitem{Lehners:2015mra}
 J.~L.~Lehners and E.~Wilson-Ewing,
 JCAP {\bf 1510} (2015) no.10, 038
 [arXiv:1507.08112 [astro-ph.CO]].




\bibitem{Mielczarek:2010ga}
 J.~Mielczarek, M.~Kamionka, A.~Kurek and M.~Szydlowski,
 JCAP {\bf 1007} (2010) 004
 [arXiv:1005.0814 [gr-qc]].



\bibitem{Lehners:2013cka}
 J.~L.~Lehners and P.~J.~Steinhardt,
 Phys.\ Rev.\ D {\bf 87} (2013) no.12, 123533
 [arXiv:1304.3122 [astro-ph.CO]].




\bibitem{Cai:2014xxa}
 Y.~F.~Cai, J.~Quintin, E.~N.~Saridakis and E.~Wilson-Ewing,
 JCAP {\bf 1407} (2014) 033
 [arXiv:1404.4364 [astro-ph.CO]].




\bibitem{Cai:2007qw}
 Y.~F.~Cai, T.~Qiu, Y.~S.~Piao, M.~Li and X.~Zhang,
 JHEP {\bf 0710} (2007) 071
 [arXiv:0704.1090 [gr-qc]].




\bibitem{Cai:2010zma}
 Y.~F.~Cai and E.~N.~Saridakis,
 Class.\ Quant.\ Grav.\ {\bf 28} (2011) 035010
 [arXiv:1007.3204 [astro-ph.CO]].

\bibitem{Avelino:2012ue}
 P.~P.~Avelino and R.~Z.~Ferreira,
 Phys.\ Rev.\ D {\bf 86} (2012) 041501
 [arXiv:1205.6676 [astro-ph.CO]].


\bibitem{Barrow:2004ad}
 J.~D.~Barrow, D.~Kimberly and J.~Magueijo,
 Class.\ Quant.\ Grav.\ {\bf 21} (2004) 4289
 [astro-ph/0406369].

\bibitem{Haro:2015zda}
 J.~Haro and E.~Elizalde,
 JCAP {\bf 1510} (2015) no.10, 028
 [arXiv:1505.07948 [gr-qc]].

\bibitem{Elizalde:2014uba}
 E.~Elizalde, J.~Haro and S.~D.~Odintsov,
 Phys.\ Rev.\ D {\bf 91} (2015) no.6, 063522
 [arXiv:1411.3475 [gr-qc]].
 
\bibitem{tp}
A. Das, D. Maity, T. Paul and S. SenGupta ; Eur.Phys.J. C77 (2017) no.12, 813 [arXiv:1706.00950]



\bibitem{Laguna:2006wr}
 P.~Laguna,
 Phys.\ Rev.\ D {\bf 75} (2007) 024033
 [gr-qc/0608117].



\bibitem{Corichi:2007am}
 A.~Corichi and P.~Singh,
 Phys.\ Rev.\ Lett.\ {\bf 100} (2008) 161302
 [arXiv:0710.4543 [gr-qc]].


\bibitem{Bojowald:2008pu}
 M.~Bojowald,
 Gen.\ Rel.\ Grav.\ {\bf 40} (2008) 2659
 [arXiv:0801.4001 [gr-qc]].




\bibitem{Singh:2006im}
 P.~Singh, K.~Vandersloot and G.~V.~Vereshchagin,
 Phys.\ Rev.\ D {\bf 74} (2006) 043510
 [gr-qc/0606032].





\bibitem{Date:2004fj}
 G.~Date and G.~M.~Hossain,
 Phys.\ Rev.\ Lett.\ {\bf 94} (2005) 011302
 [gr-qc/0407074].



\bibitem{deHaro:2012xj}
 J.~de Haro,
 JCAP {\bf 1211} (2012) 037
 [arXiv:1207.3621 [gr-qc]].





\bibitem{Cianfrani:2010ji}
 F.~Cianfrani and G.~Montani,
 Phys.\ Rev.\ D {\bf 82} (2010) 021501
 [arXiv:1006.1814 [gr-qc]].




\bibitem{Cai:2014zga}
 Y.~F.~Cai and E.~Wilson-Ewing,
 JCAP {\bf 1403} (2014) 026
 [arXiv:1402.3009 [gr-qc]].




\bibitem{Mielczarek:2008zz}
 J.~Mielczarek and M.~Szydlowski,
 Phys.\ Rev.\ D {\bf 77} (2008) 124008
 [arXiv:0801.1073 [gr-qc]].






\bibitem{Mielczarek:2008zv}
 J.~Mielczarek, T.~Stachowiak and M.~Szydlowski,
 Phys.\ Rev.\ D {\bf 77} (2008) 123506
 [arXiv:0801.0502 [gr-qc]].





\bibitem{Diener:2014mia}
 P.~Diener, B.~Gupt and P.~Singh,
 Class.\ Quant.\ Grav.\ {\bf 31} (2014) 105015
 [arXiv:1402.6613 [gr-qc]].





\bibitem{Haro:2015oqa}
 J.~Haro, A.~N.~Makarenko, A.~N.~Myagky, S.~D.~Odintsov and V.~K.~Oikonomou,
 Phys.\ Rev.\ D {\bf 92} (2015) no.12, 124026
 [arXiv:1506.08273 [gr-qc]].




\bibitem{Zhang:2011qq}
 X.~Zhang and Y.~Ma,
 Phys.\ Rev.\ D {\bf 84} (2011) 064040
 [arXiv:1107.4921 [gr-qc]].






\bibitem{Zhang:2011vi}
 X.~Zhang and Y.~Ma,
 Phys.\ Rev.\ Lett.\ {\bf 106} (2011) 171301
 [arXiv:1101.1752 [gr-qc]].



\bibitem{Cai:2014jla}
 Y.~F.~Cai and E.~Wilson-Ewing,
 JCAP {\bf 1503} (2015) no.03, 006
 [arXiv:1412.2914 [gr-qc]].

\bibitem{WilsonEwing:2012pu}
 E.~Wilson-Ewing,
 JCAP {\bf 1303} (2013) 026
 [arXiv:1211.6269 [gr-qc]].
 
 
\bibitem{Cai:2011tc}
  Y.~F.~Cai, S.~H.~Chen, J.~B.~Dent, S.~Dutta and E.~N.~Saridakis,
  Class.\ Quant.\ Grav.\  {\bf 28} (2011) 215011
  doi:10.1088/0264-9381/28/21/215011
  [arXiv:1104.4349 [astro-ph.CO]].






\bibitem{Finelli:2001sr}
 F.~Finelli and R.~Brandenberger,
 Phys.\ Rev.\ D {\bf 65} (2002) 103522
 [hep-th/0112249].




\bibitem{Cai:2011ci}
Y.~F.~Cai, R.~Brandenberger and X.~Zhang,
Phys.\ Lett.\ B {\bf 703} (2011) 25
[arXiv:1105.4286 [hep-th]].


\bibitem{Haro:2015zta}
 J.~Haro and J.~Amoros,
 PoS FFP {\bf 14} (2016) 163
 [arXiv:1501.06270 [gr-qc]].


\bibitem{Cai:2011zx}
Y.~F.~Cai, R.~Brandenberger and X.~Zhang,
JCAP {\bf 1103} (2011) 003
[arXiv:1101.0822 [hep-th]].







\bibitem{Haro:2014wha}
 J.~Haro and J.~Amoros,
 JCAP {\bf 1412} (2014) no.12, 031
 [arXiv:1406.0369 [gr-qc]].


\bibitem{Brandenberger:2009yt}
 R.~Brandenberger,
 Phys.\ Rev.\ D {\bf 80} (2009) 043516
 [arXiv:0904.2835 [hep-th]].

\bibitem{deHaro:2014kxa}
 J.~de Haro and J.~Amoros,
 JCAP {\bf 1408} (2014) 025
 [arXiv:1403.6396 [gr-qc]].


\bibitem{Odintsov:2014gea}
 S.~D.~Odintsov and V.~K.~Oikonomou,
 Phys.\ Rev.\ D {\bf 90} (2014) no.12, 124083
 [arXiv:1410.8183 [gr-qc]].



\bibitem{Qiu:2010ch}
 T.~Qiu and K.~C.~Yang,
 JCAP {\bf 1011} (2010) 012
 [arXiv:1007.2571 [astro-ph.CO]].



\bibitem{Oikonomou:2014jua}
 V.~K.~Oikonomou,
 Gen.\ Rel.\ Grav.\ {\bf 47} (2015) no.10, 126
 [arXiv:1412.8195 [gr-qc]].







\bibitem{Bamba:2012ka}
 K.~Bamba, J.~de Haro and S.~D.~Odintsov,
 JCAP {\bf 1302} (2013) 008
 [arXiv:1211.2968 [gr-qc]].
 
 
 
\bibitem{tp_bouncing1}
  S.~Nojiri, S.~D.~Odintsov, V.~K.~Oikonomou and T.~Paul,
  arXiv:1910.03546 [gr-qc].
  
  
\bibitem{Das:2017htt}
  A.~Das, H.~Mukherjee, T.~Paul and S.~SenGupta,
  Eur.\ Phys.\ J.\ C {\bf 78} (2018) no.2,  108
  doi:10.1140/epjc/s10052-018-5603-9
  [arXiv:1701.01571 [hep-th]].
  
  
\bibitem{Elizalde:2018rmz}
  E.~Elizalde, S.~D.~Odintsov, T.~Paul and D.~Sáez-Chillón Gómez,
  Phys.\ Rev.\ D {\bf 99} (2019) no.6,  063506
  doi:10.1103/PhysRevD.99.063506
  [arXiv:1811.02960 [gr-qc]].
  
  
  
\bibitem{Elizalde:2018now}
  E.~Elizalde, S.~D.~Odintsov, V.~K.~Oikonomou and T.~Paul,
  JCAP {\bf 1902} (2019) 017
  doi:10.1088/1475-7516/2019/02/017
  [arXiv:1810.07711 [gr-qc]].
 
 
 
 
 
 \bibitem{haro1}
  J.~de Haro,
  EPL {\bf 107} (2014) no.2,  29001
  doi:10.1209/0295-5075/107/29001
  [arXiv:1403.4529 [gr-qc]].
  
  
\bibitem{Capozziello:2011et}
  S.~Capozziello and M.~De Laurentis,
  Phys.\ Rept.\  {\bf 509} (2011) 167
  doi:10.1016/j.physrep.2011.09.003
  [arXiv:1108.6266 [gr-qc]].
  
  
\bibitem{Nojiri:2010wj}
  S.~Nojiri and S.~D.~Odintsov,
  Phys.\ Rept.\  {\bf 505} (2011) 59
  doi:10.1016/j.physrep.2011.04.001
  [arXiv:1011.0544 [gr-qc]].
  
  
\bibitem{Nojiri:2017ncd}
  S.~Nojiri, S.~D.~Odintsov and V.~K.~Oikonomou,
  Phys.\ Rept.\  {\bf 692} (2017) 1
  doi:10.1016/j.physrep.2017.06.001
  [arXiv:1705.11098 [gr-qc]].
 



\bibitem{Nojiri:2017ygt}
  S.~Nojiri, S.~D.~Odintsov and V.~K.~Oikonomou,
  Phys.\ Lett.\ B {\bf 775} (2017) 44
  doi:10.1016/j.physletb.2017.10.045
  [arXiv:1710.07838 [gr-qc]].
  
  
\bibitem{Nojiri:2006ri}
  S.~Nojiri and S.~D.~Odintsov,
  eConf C {\bf 0602061} (2006) 06
   [Int.\ J.\ Geom.\ Meth.\ Mod.\ Phys.\  {\bf 4} (2007) 115]
  doi:10.1142/S0219887807001928
  [hep-th/0601213].







\bibitem{hwang1}
J. c. Hwang and H. Noh, Phys. Rev. D 71 (2005) 063536
[gr-qc/0412126].

\bibitem{hwang2}
H. Noh and J. c. Hwang, Phys. Lett. B 515 (2001) 231
[astro-ph/0107069].

\bibitem{hwang3}
J. c. Hwang and H. Noh, Phys. Rev. D 66 (2002) 084009
[hep-th/0206100].


\bibitem{Akrami:2018odb}
Y.~Akrami {\it et al.} [Planck Collaboration],
arXiv:1807.06211 [astro-ph.CO].


\bibitem{Perlmutter:1998np}
  S.~Perlmutter {\it et al.} [Supernova Cosmology Project Collaboration],
  Astrophys.\ J.\  {\bf 517} (1999) 565
  doi:10.1086/307221
  [astro-ph/9812133].
  
\bibitem{Riess:1998cb}
  A.~G.~Riess {\it et al.} [Supernova Search Team],
  Astron.\ J.\  {\bf 116} (1998) 1009
  doi:10.1086/300499
  [astro-ph/9805201].



















\end{thebibliography}
\end{document}